\documentclass[pdflatex,sn-mathphys-num]{sn-jnl}

\usepackage{graphicx}%
\usepackage{multirow}%
\usepackage{amsmath,amssymb,amsfonts}%
\usepackage{amsthm}%
\usepackage{mathrsfs}%
\usepackage[title]{appendix}%
\usepackage{xcolor}%
\usepackage{textcomp}%
\usepackage{manyfoot}%
\usepackage{booktabs}%
\usepackage{algorithm}%
\usepackage{algorithmicx}%
\usepackage{algpseudocode}%
\usepackage{listings}%
\usepackage{adjustbox}%
\usepackage{makecell}%
\usepackage{subcaption}%

\theoremstyle{thmstyleone}%
\theoremstyle{thmstyletwo}%

\theoremstyle{thmstylethree}%

\begin{document}

\title{PenTiDef: Decentralized Federated Intrusion Detection System with Differential Privacy and Latent-Space Defense via Blockchain Coordination in IIoT}


\author[1,2,3]{\sur{Phan The} \fnm{Duy}}\email{duypt@uit.edu.vn}

\author[1,2,3]{\sur{Nghi Hoang} \fnm{Khoa}}\email{khoanh@uit.edu.vn}

\author[1,2,3]{\sur{Nguyen Tran Anh} \fnm{Quan}}\email{20521793@gm.uit.edu.vn}

\author[1,2,3]{\sur{Luong Ha} \fnm{Tien}}\email{20520802@gm.uit.edu.vn}

\author[1,2,3]{\sur{Ngo Duc Hoang} \fnm{Son}}\email{sonndh@uit.edu.vn}

\author*[1,2,3]{\fnm{Van-Hau} \sur{Pham}}\email{haupv@uit.edu.vn}

\affil[1]{\orgdiv{Information Security Lab}, \orgname{University of Information Technology}, \orgaddress{\city{Ho Chi Minh City},  \country{Vietnam}}}

\affil[2]{\orgname{Vietnam National University}, \orgaddress{\city{Ho Chi Minh City},  \country{Vietnam}}}

\affil[3]{\orgname{VNU-HCM Information Security Center}, \orgaddress{\city{Ho Chi Minh City},  \country{Vietnam}}}

\abstract{
This paper proposes PenTiDef, a fully decentralized, privacy-preserving, and poisoning-resilient framework for decentralized federated IDS (DFL-IDS). PenTiDef synergistically integrates three key components: (i) client-side Distributed Differential Privacy (DDP) with stochastic Gaussian noise to protect gradient leakage, (ii) a lightweight latent-space defense module that extracts and compresses penultimate-layer representations (PLRs) into stable Latent Semantic Representations (LSRs) via AutoEncoder, followed by Centered Kernel Alignment (CKA) and K-Means clustering for robust malicious update detection without auxiliary datasets, and (iii) a permissioned blockchain layer with smart contracts that orchestrates on-chain validation, secure FedAvg aggregation, and immutable auditability, eliminating any central server. Extensive experiments on CIC-IDS2018 and Edge-IIoTSet under both IID and realistic non-IID settings, with adversary ratios up to 40\%, demonstrate that PenTiDef consistently outperforms state-of-the-art baselines (FLARE and FedCC) in detection accuracy and F1-score while maintaining lower training overhead. By jointly addressing privacy, robustness, and decentralization in a unified secure aggregation protocol, PenTiDef provides a practical and scalable solution for trustworthy collaborative intrusion detection in heterogeneous, adversarial IIoT environments.
}


\maketitle

\section{Introduction}
\label{sec:intro}

The transition toward Industry 4.0 has dramatically increased the connectivity of production systems by integrating IIoT components such as sensors, actuators, industrial gateways, communication protocols, and edge-side computing resources \cite{alabadi2022industrial, 2026chen_iiot_survey}. These interconnected infrastructures enable real-time monitoring, predictive maintenance, operational efficiency, and data-driven decision-making across industrial sectors \cite{afrin2025industrial}. In this context, Intrusion Detection Systems (IDS) play a critical role in safeguarding industrial networks by monitoring traffic and identifying abnormal behaviors across multiple layers, from field-level device communications to gateway-aggregated flows and edge-side monitoring streams \cite{Rezaei2026_ids_survey, hakeem2025advancing}.

With the advancement of Artificial Intelligence (AI), Machine Learning (ML) techniques have been widely adopted in IDS to learn complex attack patterns from large-scale operational and network data generated by modern industrial systems \cite{Nkoro2026_ids_AI_survey}. However, training effective ML-based IDS models requires extensive data from diverse sources, including multiple attack types, machine-to-machine communications, sensor telemetry, and gateway traffic. Data sharing is severely restricted due to privacy and confidentiality concerns. Federated Learning (FL) has emerged as a promising paradigm to address this tension by enabling collaborative model training without exchanging raw data \cite{agrawal2021federated, khan2021federated, vahabi2025federated}.

Despite its advantages, FL remains vulnerable to malicious participants. Attackers can launch poisoning attacks by uploading manipulated model updates, thereby degrading global model performance and compromising the reliability of intrusion detection in industrial operations. Such threats are particularly concerning in decentralized settings. Consequently, integrating blockchain with FL has become a relevant direction to enhance trust, auditability, and resilience in decentralized IIoT security infrastructures \cite{issa2023blockchain}. Recent studies \cite{hu2021gfl, BCFL_survey, 2026_quan_FL_cps_survey} have demonstrated the potential of blockchain as a decentralized, tamper-proof ledger for model updates and coordination among multiple production sites or edge domains. Its immutability and transparency help prevent tampering while eliminating single points of failure. Nevertheless, FL still faces significant privacy risks \cite{fu2026differentially, Zuo2026_privacyrisk_FL_sur}. Model updates can leak sensitive information about local industrial traffic and device behavior through gradient inversion or membership inference attacks.

To mitigate these privacy threats, various cryptographic approaches such as Homomorphic Encryption (HE) \cite{HE_privacy} and Secure Multi-Party Computation (SMPC) \cite{smpc_privacy, smpc, secureml} have been proposed. However, these methods incur high computational and communication overhead, making them impractical for resource-constrained IIoT environments. Differential Privacy (DP) offers a more scalable alternative by adding calibrated noise to model updates \cite{Privacy_AntiPoi}. In parallel, poisoning attacks remain a major barrier to DFL adoption in IIoT \cite{Privacy_AntiPoi, Nowroozi2025_FL_under_attack_sur, uddin2025systematic}. Most existing defenses rely on anomaly detection or robust aggregation \cite{li2021lomar, afl_cs, 10018378, li2022robust, li2021byzantine, 9721118}, yet they often suffer from high computational cost, the need for prior knowledge of malicious participants, or poor performance under non-IID data distributions typical in industrial settings \cite{zhu2021federated, zhao2021federated}.

Recent latent-space-based methods, including FedCC \cite{fedcc}, FLARE \cite{wang2022flare}, and Fed-LSAE \cite{luong2024fed}, have shown promising results by comparing penultimate layer representations (PLRs). While FLARE requires auxiliary datasets and FedCC suffers from PLR instability in non-IID scenarios, Fed-LSAE improves stability through AutoEncoder compression. However, these approaches remain limited to centralized or semi-centralized architectures and lack explicit privacy guarantees or fully decentralized coordination.

In this paper, we propose PenTiDef, a novel blockchain-orchestrated, privacy-preserving, and poisoning-resilient framework specifically designed for decentralized federated intrusion detection in heterogeneous Industrial IoT (IIoT) environments. Unlike prior works that address individual aspects of the problem in isolation, PenTiDef presents the first holistic architecture that simultaneously tackles centralized single-point-of-failure, gradient leakage, and poisoning attacks under severe non-IID data distributions—challenges that are particularly critical in multi-site industrial deployments.

The key contributions of this work are as follows:

\begin{itemize}
    \item A fully decentralized FL-IDS coordination architecture that eliminates any central aggregation server by leveraging a permissioned blockchain (Hyperledger Fabric) with smart contracts. This design enforces transparent model validation, immutable auditability, and incentive-compatible reward/punishment mechanisms across untrusted administrative domains, significantly enhancing system fault tolerance and trust without sacrificing performance.

    \item A client-level Distributed Differential Privacy (DDP) mechanism with stochastic Gaussian noise injection that provides formal $(\epsilon, \delta)$-privacy guarantees on local model updates while maintaining high model utility. This is the first integration of DDP into a latent-space poisoning defense pipeline for DFL-IDS, effectively mitigating gradient-based inference attacks in resource-constrained IIoT settings.

    \item A latent-space poisoning defense module tailored for decentralized IIoT environments. It integrates AutoEncoder-compressed Latent Semantic Representations (LSRs) from penultimate layers with CKA and unsupervised clustering, specifically hardened for severe non-IID industrial traffic. Unlike FedCC (PLR instability) and Fed-LSAE (centralized), the module operates without auxiliary data or adversary knowledge and is tightly coupled with on-chain DDP updates and blockchain validation for robust, privacy-preserving detection.

    \item A synergistic integration of the above components into a lightweight, end-to-end pipeline where DDP-perturbed updates are validated on-chain in the latent space before secure aggregation. This integration achieves superior robustness against both targeted and untargeted poisoning attacks (label-flipping, weight-scaling, Krum/Med, backdoor, and GAN-based) while delivering lower training overhead and better scalability than state-of-the-art defenses.

    \item Extensive empirical validation on two large-scale IDS benchmarks (CIC-IDS2018 and Edge-IIoTSet) under both IID and realistic non-IID settings, with adversary ratios up to 40\%. Results demonstrate that PenTiDef consistently outperforms FLARE, FedCC, and other baselines in detection accuracy, F1-score, and convergence speed, while providing strong privacy protection and blockchain-level auditability.
\end{itemize}

By jointly addressing decentralization, privacy, and robustness in a single cohesive framework, PenTiDef establishes a new benchmark for deploying trustworthy federated intrusion detection systems in adversarial and heterogeneous IIoT infrastructures.

This article is organized as follows in the remaining sections. In Section \ref{sect_related}, relevant papers on privacy attacks and poisoning attacks against FL-based models, and their defenses are discussed. The threat model is presented in Section \ref{Sec:threatmodel} to determine the assumption about attackers in our works. Section \ref{sec:method} then discusses the approach and threat model. Next, in Section \ref{sec:researchquestion}, we outline the research questions and experimental scenarios to benchmark the framework. The experimental setups, dataset and evaluation metrics are presented in Section \ref{sec:implementation}. Following that, Section \ref{sec:results} provides the experimental results and analysis of the PenTiDef performance. The discussion on achievement and limitation of our work is mentioned in Section \ref{sec:discus}. In Section \ref{sec:conclusion}, we finally conclude the paper.

\section{Related work} \label{sect_related}

\subsection{Privacy Attacks in the Context of FL}

Although FL avoids direct sharing of raw data among participants, it remains susceptible to privacy leakage through exchanged model updates. Prior studies \cite{Privacy_exploiting, HE, DeepLeak} have demonstrated that gradients and consecutive model snapshots can inadvertently expose sensitive information about local training data. This vulnerability stems from the fact that deep learning models often encode latent representations beyond the primary task objective.

Adversaries can exploit these gradients to perform various inference attacks, aiming to recover private information such as class representatives, membership status, sample attributes, and even training inputs or labels. One notable example is the GAN-based attack, where malicious clients leverage Generative Adversarial Networks to synthesize prototypical data from other participants. Another common threat is membership inference, which seeks to determine whether a specific data point was part of a client’s training set. These attacks can be executed passively—by observing model updates or actively, through deliberate manipulation of the training process to amplify information leakage \cite{Privacy_AntiPoi}.

These findings underscore the importance of incorporating robust gradient protection mechanisms in FL, especially in privacy-sensitive applications such as healthcare, finance, and smart environments.

\begin{table*}[!t]
    \centering
    \caption{Comparison between DP categorizations}
    \begin{tabular}{lccc}
        \toprule
        \textbf{DP type} & \textbf{Trusted aggregator} & \textbf{Adding noise by} & \textbf{Privacy Guarantee} \\
        \toprule
        CDP  & Yes & Aggregator & Aggregated value \\
        LDP  & No  & User       & Locally released value \\
        DDP  & No  & User       & Aggregated value \\
        \bottomrule
    \end{tabular}
    \label{tab:DP_types}
\end{table*}

\begin{table*}[!b]
    \centering
     \caption{Comparison of studies on defending against poisoning attacks in FL}
    \label{tab:compares_defending_poison}
    \begin{adjustbox}{width=\textwidth}
    \begin{tabular}{cccccccc}
    
    \toprule
    \textbf{Work} &
    \textbf{\makecell{Attack\\strategies}} &
    \textbf{\makecell{FL\\approach}} &
    \textbf{\makecell{Feature}} &
    \textbf{\makecell{Detection\\mechanism}} &
    \textbf{\makecell{Privacy\\mechanism}} &
    \textbf{Non-IID} &
    \textbf{Dataset}\\
    \toprule
    \makecell{ShieldFL\\\cite{shieldfl}} &
    \makecell[l]{+Untargeted:\\\hspace{1em}- Label flipping\\+Targeted:\\\hspace{1em}- Label flipping} &
    CFL &
    \makecell{Local\\model\\gradient} &
    \makecell{Cosine\\similarity} &
    HE &
    × &
    \makecell{MNIST\\KDDCup\\Amazon} \\

    \midrule
    \makecell{FLARE\\\cite{wang2022flare}} &
    \makecell[l]{+Untargeted:\\\hspace{1em}- Attack-Krum-Untargeted\\\hspace{1em}- Attack-TM-Untargeted\\+Targeted:\\\hspace{1em}- Attack-Krum-Backdoor\\\hspace{1em}- Attack-Coomed-Backdoor} &
    CFL &
    PLR &
    \makecell{Maximum\\mean\\discrepancy} &
    - &
    - &
    \makecell{fMNIST\\CIFAR-10\\Kather} \\

    \midrule
    \makecell{FedCC\\\cite{fedcc}}           &
    \makecell[l]{+Untargeted:\\\hspace{1em}- Untargeted-Krum\\\hspace{1em}- Untargeted-Med\\+Targeted:\\\hspace{1em}- Backdoor} &
    CFL &
    PLR &
    \makecell{CKA\\similarity\\with\\Clustering} &
    - &
    - &
    \makecell{fMNIST\\CIFAR10\\CIFAR100}\\

    \midrule
    \makecell{Liu et al.\\\cite{liutblockflvec}} &
    \makecell[l]{+Untargeted:\\\hspace{1em}- Model poisoning\\\hspace{1em}- Plaintext attack\\\hspace{1em}- Untargeted-Med\\+Targeted:\\\hspace{1em}- Backdoor injection\\\hspace{1em}- GAN
    \\} &
    DFL &
    \makecell{Model\\parameters} &
    \makecell{Low-accuracy\\detection in\\(m,n)\\threshold\\aggregated\\models} &
    \makecell{Secret Sharing\\Differential\\Privacy\\Blockchain} &
    - &
    \makecell{KDDCup99}\\

    \midrule
    \makecell{DeepBlockIoTNet\\\cite{rathoreblocldisdl}} &
    \makecell[l]{+Untargeted:\\\hspace{1em}- Model poisoning\\\hspace{1em}- Untargeted-Med\\+Targeted:\\\hspace{1em}- Adversarial injection\\\hspace{1em}- Server poisoning\\} &
    \makecell{DFL} &
    \makecell{Encrypted\\gradient,\\signature} &
    \makecell{Voting,\\Model\\accuracy\\deviation} &
    \makecell{Public-Key\\Encryption,\\Digital Signature,\\Blockchain} &
    - &
    \makecell{COCO}\\

    \midrule
    \makecell{Fed-LSAE\\\cite{luong2024fed}} &
    \makecell[l]{+Untargeted:\\\hspace{1em}- Label flipping\\\hspace{1em}- Weight scaling\\\hspace{1em}- Untargeted-Med\\+Targeted:\\\hspace{1em}- GAN\\} &
    CFL &
    \makecell{Computed\\LSR from\\PLR via\\AE} &
    \makecell{CKA\\similarity\\with\\Clustering\\of LSRs} &
    - &
    × &
    \makecell{CIC-ToN-IoT\\N-BaIoT}\\
    
    \midrule
    \textbf{\makecell{PentiDef\\(Our)}}             &
    \makecell[l]{+Untargeted:\\\hspace{1em}- Label flipping\\\hspace{1em}- Weight scaling\\\hspace{1em}- Untargeted-Krum\\\hspace{1em}- Untargeted-Med\\+Targeted:\\\hspace{1em}- Backdoor\\\hspace{1em}- GAN} &
    DFL &
    \makecell{Computed\\LSR from\\PLR via\\AE} &
    \makecell{CKA\\similarity\\with\\Clustering\\of LSRs} &
    DDP &
    × &
    \makecell{Edge-IIoTSet\\CIC-IDS2018}\\

    \bottomrule
    \end{tabular}
    \end{adjustbox}
\end{table*}

\subsection{Defense Mechanisms Against Privacy Attacks in FL}

Preserving privacy in FL poses distinct challenges due to decentralized data, statistical heterogeneity, and limited communication reliability. Among the major privacy-preserving strategies proposed, three prominent approaches are Homomorphic Encryption (HE), Secure Multi-Party Computation (SMPC), and Differential Privacy (DP) \cite{Privacy_AntiPoi, shi2023privacy}.

HE enables computations on encrypted data without decryption and is categorized into Fully, Somewhat, and Partially Homomorphic Encryption. While Fully HE supports arbitrary computations, it is computationally intensive and impractical for real-time FL scenarios. In contrast, Partially HE schemes—such as RSA, El Gamal, and Paillier—are more efficient but limited in function \cite{Privacy_AntiPoi}. Despite its theoretical appeal, HE incurs substantial memory and runtime overhead, hindering scalability in decentralized FL (DFL).

SMPC allows multiple parties to collaboratively compute functions over private inputs without revealing them. For instance, Şahinbaş and Catak \cite{sahinbas2023secure} proposed an SMPC-based framework for FL in healthcare IoT, enabling secure model training while preserving data confidentiality. Similarly, SecureML by Mohassel and Zhang \cite{mohassel2017secureml} enables distributed model training via secret sharing between non-colluding servers. However, both approaches suffer from high computational and communication costs, making them less suitable for large-scale or resource-constrained environments.

In contrast, DP provides a lightweight yet effective privacy mechanism by introducing calibrated noise into data or model updates. DP is commonly categorized into three variants: Centralized DP (CDP), Local DP (LDP), and Distributed DP (DDP), as summarized in Table~\ref{tab:DP_types}. CDP depends on a trusted aggregator to inject noise after model update collection, which introduces a central point of failure and regulatory concerns. LDP removes the need for a trusted server by having clients perturb data locally; however, it often compromises model accuracy in high-dimensional settings.

DDP overcomes the limitations of CDP and LDP by distributing noise addition across participants using stable distributions, such as Gaussian \cite{liu2018generalized} and Binomial \cite{10.1007/978-3-030-17653-2_13}. In the context of DFL, DDP enhances privacy without requiring centralized trust, reduces the impact of noise on model utility, and supports scalability. These characteristics make DDP particularly suitable for privacy-preserving learning in adversarial and heterogeneous FL environments.

\subsection{Poisoning Attacks and Defense Mechanisms in FL}

Despite its privacy advantages, FL remains vulnerable to poisoning attacks, which aim to degrade global model performance through manipulated client updates. These attacks are typically categorized as data poisoning—where adversaries inject mislabeled or malicious samples into their local datasets—and model poisoning, which involves directly altering model parameters to influence aggregation. Model poisoning is generally more effective, as it circumvents the need for data manipulation and directly impacts global convergence.

Recent studies have explored various attack strategies. Zhang et al. introduced GAN-based attacks \cite{GAN_nets}, where adversaries use the global model as a discriminator to generate malicious samples that appear benign. Building on this, PoisonGAN \cite{PoisonGAN} targets FL in edge computing by injecting adversarial updates crafted via generative models, demonstrating high success rates in both backdoor and label-flipping scenarios. Similarly, the PoisonedFL framework \cite{PoisonedFL} utilizes multi-round consistency among malicious clients to evade detection, outperforming multiple state-of-the-art defenses in targeted settings.

In response, several defense mechanisms have been proposed. ShieldFL \cite{shieldfl} employs homomorphic encryption with cosine similarity scoring and a Byzantine-tolerant aggregation strategy to mitigate poisoning risks. While effective, its cryptographic overhead limits scalability in large-scale deployments.

Recent approaches have focused on leveraging latent space representations (LSR) for model-level anomaly detection. FLARE \cite{wang2022flare} extracts penultimate-layer representations (PLR) using an auxiliary dataset to compute trust scores, filtering out low-trust models. However, its reliance on external data introduces potential privacy leakage and reduces generalizability. FedCC \cite{fedcc} addresses this by using the CKA algorithm to compare local and global PLRs without auxiliary data, clustering models via K-Means. Although FedCC improves resilience against model poisoning, it may suffer from PLR instability in non-IID settings.

To overcome these limitations, Fed-LSAE \cite{luong2024fed} introduces an AutoEncoder-based approach that compresses PLRs into stable LSRs without auxiliary data or prior knowledge. By distinguishing malicious from benign models based on latent characteristics, Fed-LSAE improves detection accuracy in non-IID environments and reduces the risk of false positives.

\subsection{Positioning and Advancements over Prior Works}

While substantial progress has been made in defending federated learning (FL) systems against poisoning attacks, existing solutions remain inadequate for real-world decentralized FL-IDS (DFL-IDS) deployments in heterogeneous Industrial IoT (IIoT) environments. Centralized approaches such as FLARE~\cite{wang2022flare} rely on auxiliary datasets for penultimate-layer representation (PLR) trust scoring, violating strict privacy constraints and limiting generalizability. FedCC~\cite{fedcc} mitigates the auxiliary data requirement but suffers from PLR instability under non-IID distributions and still assumes a trusted central aggregator. Although Fed-LSAE~\cite{luong2024fed} advances latent-space inspection via AutoEncoder compression, it inherits the same centralized architecture and lacks explicit privacy guarantees or Byzantine-resilient coordination.

PenTiDef advances the state of the art by presenting the first comprehensive framework that simultaneously achieves:
\begin{itemize}
    \item \textbf{Fully decentralized coordination} without any single point of aggregation or trust, enabled by a permissioned blockchain (Hyperledger Fabric) layer with smart-contract-driven model validation, reward/punishment, and immutable audit trails;
    \item \textbf{Strong client-level privacy} through Distributed Differential Privacy (DDP) with stochastic Gaussian noise, providing formal $(\epsilon,\delta)$-guarantees while preserving model utility in high-dimensional IIoT traffic data;
    \item \textbf{Robust latent-space poisoning detection} that extends and stabilizes Fed-LSAE-style representations by integrating AutoEncoder-compressed Latent Semantic Representations (LSRs) with Centered Kernel Alignment (CKA) and unsupervised clustering—explicitly designed to handle severe non-IID distributions common in multi-site industrial deployments.
\end{itemize}

Crucially, these components are not merely juxtaposed but synergistically integrated: DDP-perturbed updates are validated in the latent space on-chain, enabling secure FedAvg aggregation of only benign contributions without auxiliary data or prior knowledge of adversary numbers. This design delivers superior detection performance against both targeted and untargeted poisoning attacks (including adaptive Krum/Med, backdoor, and GAN-based) under 10--40\% adversary ratios, while achieving lower training overhead compared to baselines.

Table~\ref{tab:compares_defending_poison} quantitatively demonstrates that PenTiDef establishes a new Pareto frontier in the joint space of \textit{privacy}, \textit{robustness}, \textit{decentralization}, and \textit{scalability} for DFL-IDS---capabilities that no prior work has simultaneously delivered in heterogeneous IIoT settings.

\section{Threat model} \label{Sec:threatmodel}
\subsection{Overview}

For this research, we assume that FL-based IDS is constructed with 20 participating collaborators, setting the number of attackers at 10\%, 20\% and 40\% (equivalent to 2, 4, and 8 attackers, respectively). This proportion ensures that the number of attackers is less than half of the total number of participants. The training consisted of 10 rounds, and the attackers executed poisoning attacks from the very first round until the last. Throughout the global model training process, the remaining participants are considered trustworthy. All trainers are required to participate in FL training, contributing to and fully updating their local models.

\subsection{Attack strategy}
\label{attack_strategy}
\subsubsection{Untargeted attack}
\begin{enumerate}
    \item \textbf{Label-flipping attack}: is a form of attack where the attacker intentionally changes the labels of training data samples from correct to incorrect, aiming to distort the model training process and reduce the model's accuracy in classifying data samples. Our ML/DL model is a binary classifier, with benign samples labeled as 0 and attack traffic samples labeled as 1. We simulate an attacker who flips all these labels to their opposite values.
    \item \textbf{Weight Scaling Attack}: This attack technique involves the attacker scaling the weights of the local model by a large factor before sending them. The goal of weight scaling is to influence the global model aggregation process, causing undesirable or abnormal changes in the final model. This can lead to a skewed or inaccurate global model, degrading system performance. To simulate this attack, we apply the formula:
    \[
    \tilde{w}_i = \lambda \cdot w_i
    \]
    where:
    \begin{itemize}
        \item \(\tilde{w}_i\) is the scaled weight of the \(i\)-th local model.
        \item \(w_i\) is the original weight of the \(i\)-th local model.
        \item \(\lambda\) is the scaling factor, typically a value greater than 1.
    \end{itemize}
    
    \item \textbf{Untargeted-Krum \cite{fedcc}}: This attack aims to degrade the performance of the global model without a specific target. The attacker sends manipulated but seemingly benign model parameters to maximize their acceptance by the Krum algorithm. Krum selects the most reliable local model based on the Euclidean distance between model parameters. However, Untargeted-Krum attack attempts to make the malicious parameters resemble benign ones to be selected, harming the global model without easy detection. 
    
    
    \item \textbf{Untargeted-Med \cite{fedcc}}: This attack manipulates model parameters based on the maximum and minimum values of the model parameters so that the median value by coordinate moves in the opposite direction. 
    
    
    
    The goal of Untargeted-Med attacks is to reduce the performance of the global model without a specific target. Attackers can degrade the model's performance without detection by altering model parameters so that the median coordinate value moves in the opposite direction. This is done by changing parameter values to fall outside the range of benign values.
\end{enumerate}

\begin{figure*}[!t]
    \centering
   
    \includegraphics[width=15cm]{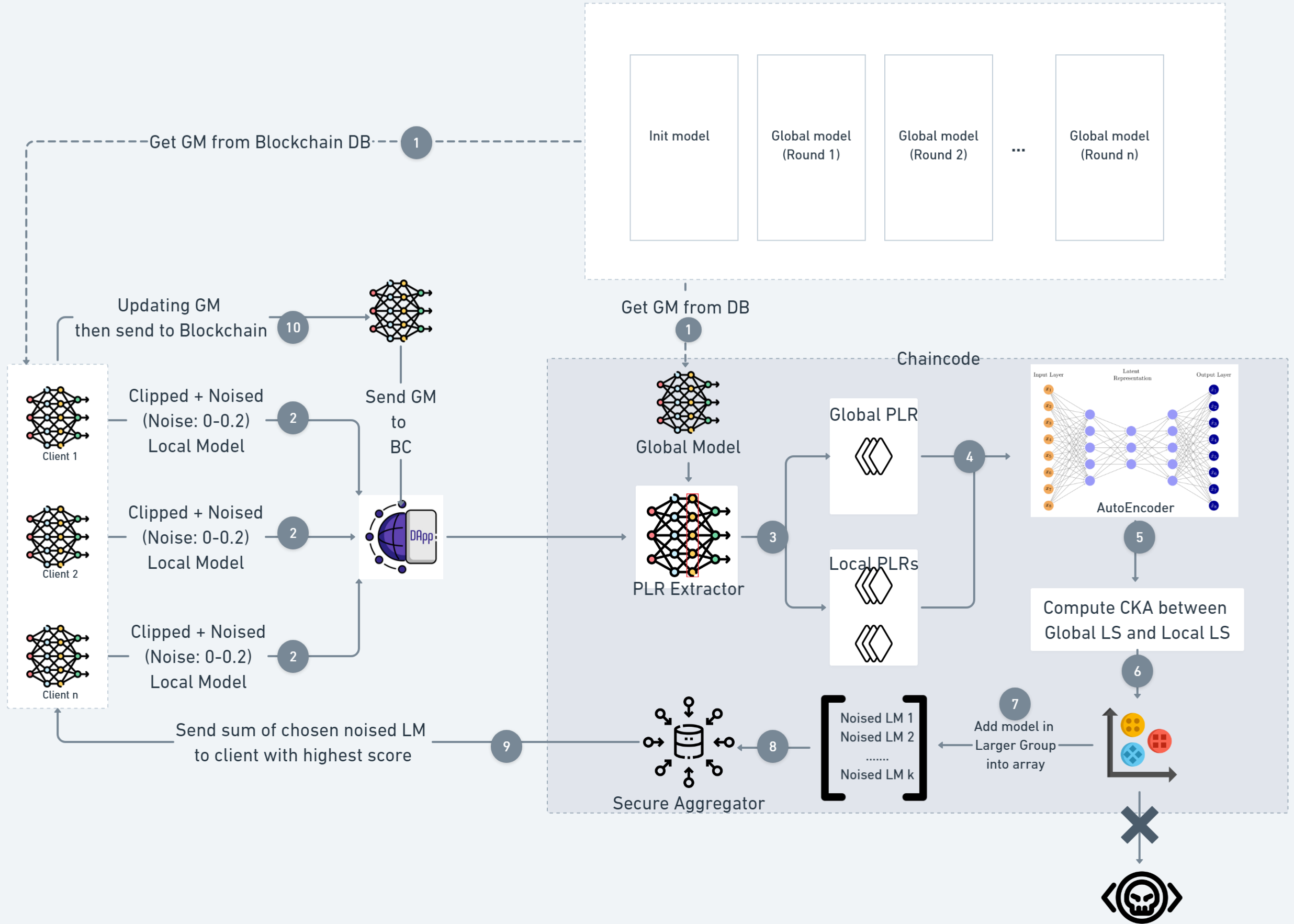}
    \caption{The PenTiDef architecture for privacy-preserving and anti-poisoning attack mechanisms in DFL-based IDS} \label{fig:PenTiDef Module}
\end{figure*}

\subsubsection{Targeted attack}
\begin{enumerate}
    \item Backdoor \cite{poisoning_survey}: A backdoor attack involves injecting a malicious model into the existing model while maintaining the accuracy of other tasks. This means that the attacked model still performs well on the original tasks, but behaves undesirably when encountering a specific trigger. In our simulation, the backdoor trigger is added to the training data by checking the average value of each sample and changing its label if the average value exceeds a threshold. This can cause the global model to behave unexpectedly when it encounters samples with the trigger during prediction.
    \item GAN poisoning attack \cite{GAN}: This involves using a type of artificial neural network to generate new data that resemble real data. A GAN consists of two sub-networks: a \textbf{Generator} and a \textbf{Discriminator}. The generator creates new data from an input dataset, while the discriminator distinguishes between real and generated data. By introducing fake data, the attacker trains with their malicious local model, thereby degrading and distorting the global model's classification performance.
\end{enumerate}

\section{Proposed Architecture}
\label{sec:method}

PenTiDef is a fully decentralized, privacy-preserving, and poisoning-resilient framework for Federated Intrusion Detection Systems (FL-IDS) in heterogeneous IIoT environments. It integrates three synergistic components into a single secure model aggregation pipeline: (i) client-level Distributed Differential Privacy (DDP) for gradient protection, (ii) latent-space anomaly detection for Byzantine robustness, and (iii) blockchain-orchestrated coordination for trust and auditability---all executed without a central server or auxiliary datasets.

\subsection{Overall Design and Threat Integration}

Let $\mathcal{C} = \{c_1, \dots, c_n\}$ be the set of collaborating IIoT nodes, each holding a private local dataset $D_i$. At global round $t$, every node $c_i$ computes a local update $w_i^t \gets \text{ClientUpdate}(w^{t-1}, D_i)$. PenTiDef enforces the following secure aggregation protocol:

\[
w^t \gets \text{PenTiDefAgg}\Bigl( \bigl\{ \tilde{w}_i^t \bigr\}_{i=1}^n, \, w^{t-1} \Bigr)
\]

where $\tilde{w}_i^t = w_i^t + \mathcal{N}(0, \sigma_i^2)$ denotes the DDP-perturbed update (detailed in Section~\ref{section:ddp-noise}). The aggregation function $\text{PenTiDefAgg}$ performs on-chain validation and selective FedAvg as formalized in Algorithm~\ref{alg:PenTiDef}.

Figure~\ref{fig:PenTiDef Module} illustrates the end-to-end pipeline. Key architectural components are tightly coupled as follows:

\begin{itemize}
    \item \textbf{Blockchain Coordination Layer} (Hyperledger Fabric): Acts as both immutable ledger and active orchestrator. Smart contracts (chaincode) receive DDP-perturbed updates, invoke latent-space validation, enforce selection of benign models, and record metadata (hash pointers, $\sigma_i$, trust scores) for auditability.
    
    \item \textbf{Latent-Space Defense Module}: Extracts penultimate-layer representations (PLRs) from both global model $w^{t-1}$ and perturbed local models $\{\tilde{w}_i^t\}$. These PLRs are compressed into stable Latent Semantic Representations (LSRs) via a lightweight AutoEncoder. Centered Kernel Alignment (CKA) with unsupervised K-Means clustering then identifies and filters poisoned contributions under non-IID conditions.
    
    \item \textbf{Privacy Layer (DDP)}: Gaussian noise is injected locally before any transmission. The perturbed models are used consistently in both latent detection and secure aggregation, ensuring end-to-end privacy without trusted third parties.
\end{itemize}

This design eliminates centralized aggregation risks while guaranteeing that only benign, privacy-protected updates contribute to the global model.

\subsection{Unified PenTiDef Aggregation Protocol}

PenTiDef realizes a fully integrated secure aggregation pipeline that tightly couples client-level privacy protection, latent-space poisoning detection, and blockchain-orchestrated coordination. The complete process is formalized in Algorithm~\ref{alg:PenTiDef}.

\begin{algorithm}[!htp]
\caption{PenTiDef: Secure Decentralized Aggregation for Poisoning-Resilient DFL-IDS}
\label{alg:PenTiDef}
\begin{algorithmic}[1]
\Require Global model $W^{t-1}$, set of collaborating clients $\mathcal{C}=\{c_1,\dots,c_n\}$, AutoEncoder $AE$
\Ensure Aggregated global model $W^t$, index of most aligned client $max\_idx$

\Statex \textbf{Phase 0: Client-side Local Training and DDP Perturbation (at each $c_i$)}
\For{each client $c_i \in \mathcal{C}$ \textbf{in parallel}}
    \State $w_i^t \gets \text{ClientUpdate}(W^{t-1}, D_i)$ \quad \Comment{local training on private data $D_i$}
    \State $\tilde{w}_i^t \gets w_i^t + \mathcal{N}(0, \sigma_i^2 \mathbf{I})$ \quad \Comment{DDP Gaussian perturbation, $\sigma_i \sim U[0,0.2]$}
    \State Send $\tilde{w}_i^t$ (with signature and $\sigma_i$) to blockchain network
\EndFor

\Statex \textbf{Phase 1: On-chain Latent Representation Analysis}
\For{$i \gets 1$ to $n$}
    \State $plr_i \gets \tilde{w}_i^t[\text{penultimate layer}]$
\EndFor
\State $plr_g \gets W^{t-1}[\text{penultimate layer}]$
\State $lsr_g \gets AE.\text{Encoder}(plr_g)$
\For{$i \gets 1$ to $n$}
    \State $lsr_i \gets AE.\text{Encoder}(plr_i)$
    \State $cka_i \gets \text{CKA}(lsr_g, lsr_i)$ \quad \Comment{linear kernel, mean-centered \& L2-normalized}
\EndFor
\State $result \gets \text{KMeans}(n\_clusters=2, \{cka_i\})$
\State $benign \gets result.\text{majority group}$

\Statex \textbf{Phase 2: Secure Aggregation and Blockchain Coordination}
\State $W^t \gets \text{FedAvg}(\{ \tilde{w}_i^t \mid i \in benign \})$
\State $max\_idx \gets \arg\max_i \, cka_i$
\State Broadcast $W^t$ via smart contract to client $c_{max\_idx}$ for next round
\Return $W^t$, $max\_idx$
\end{algorithmic}
\end{algorithm}

Algorithm~\ref{alg:PenTiDef} presents the complete PenTiDef protocol. At each client (Phase 0), local model training is immediately followed by Distributed Differential Privacy (DDP) perturbation using the Gaussian mechanism with stochastic noise scale $\sigma_i \in [0,0.2]$. The resulting perturbed updates $\tilde{w}_i^t$ are used consistently in both on-chain latent-space validation (Phase 1) and secure aggregation (Phase 2). This ensures that privacy protection is enforced end-to-end: poisoned or unperturbed updates never enter the aggregation step.

The linear-kernel CKA formulation, computed after mean-centering and L2-normalization of the LSRs, provides robust structural similarity measurement that is resilient to both non-IID industrial data distributions and the injected DDP noise. Its superior separation capability compared to cosine or Euclidean metrics has been validated empirically, enabling reliable identification of malicious contributions even under severe heterogeneity and up to 40\% adversary ratios.

Smart contracts on the permissioned blockchain (Hyperledger Fabric) embed the entire validation logic of Phases 1--2, ensuring that only cryptographically signed, DDP-protected, and latent-verified updates are accepted. This unified design eliminates centralized single points of failure while guaranteeing privacy, robustness, and auditability within a single coherent protocol.

\subsection{Distributed Differential Privacy Mechanism}
\label{section:ddp-noise}

PenTiDef enforces client-side privacy through Distributed Differential Privacy (DDP) using the Gaussian mechanism. For each local update $w_i^t$ computed on private dataset $D_i$, the perturbed model is generated as

\[
\tilde{w}_i^t = w_i^t + \mathcal{N}(0, \sigma_i^2 \mathbf{I}),
\]

where the noise scale is

\[
\sigma_i = \frac{\sqrt{2 \ln(1.25/\delta)} \cdot \Delta_2}{\epsilon},
\]

with $\Delta_2$ being the $L_2$-sensitivity of the update function and $(\epsilon, \delta)$ the target privacy budget. To increase robustness against adaptive adversaries and free-riders, $\sigma_i$ is stochastically sampled from $[0, 0.2]$ independently at every round and client.

This perturbation is applied \emph{before} transmission (Phase 0 of Algorithm~\ref{alg:PenTiDef}), ensuring that both latent-space validation and secure aggregation operate exclusively on privacy-protected updates. The chosen noise range maintains model utility while providing meaningful $(\epsilon, \delta)$-guarantees. Each $\sigma_i$ is recorded as on-chain metadata for post-hoc privacy auditing without exposing raw weights.

\subsection{Blockchain-Orchestrated Trust Enforcement}

The blockchain layer serves as the trust and coordination backbone of PenTiDef, replacing any central server. We implement it using a permissioned Hyperledger Fabric network with 3 organizations and 6 peer nodes, providing fault tolerance across heterogeneous IIoT administrative domains.

Each DDP-perturbed update $\tilde{w}_i^t$ is stored off-chain (IPFS) and referenced on-ledger via a cryptographic hash $h(\tilde{w}_i^t)$ together with metadata tuple $\langle h(\tilde{w}_i^t), \texttt{meta}_i^t \rangle$, where $\texttt{meta}_i^t$ contains round ID, client certificate, $\sigma_i$, and latent trust score. Smart contracts embed the full validation logic of Algorithm~\ref{alg:PenTiDef} (PLR$\to$LSR$\to$CKA$\to$K-Means filtering) directly into the transaction pipeline:

\[
\text{VerifySig}(\mathsf{ID}_i, \tau) \ \land \ \text{LatentVerify}(\tilde{w}_i^t) \ \land \ \text{ValidateMeta}(\texttt{meta}_i^t) \ \implies \ \text{Append}(\tau, \mathcal{L}).
\]

Only verified benign updates proceed to FedAvg aggregation. The resulting global model $W^t$ is then distributed via smart contract to the client with the highest CKA score. This design delivers immutable auditability, incentive-compatible participation (through reward/penalty mechanisms), and complete elimination of centralized single points of failure.

Together with the DDP mechanism and latent-space module, the blockchain layer forms an interdependent, end-to-end secure protocol that achieves strong privacy, robustness against poisoning, and practical decentralization in adversarial IIoT environments.

\subsection{Computational, Communication, and Storage Overhead Analysis}

To evaluate the practical feasibility of PenTiDef in resource-constrained Industrial IoT (IIoT) and Cyber-Physical Systems (CPS), we analyze its computational, communication, and storage overhead based on the unified protocol in Algorithm~\ref{alg:PenTiDef}, the experimental setup with $n=20$ collaborating clients, and the CNN-based IDS model described in the implementation.

\textbf{Computational Complexity:}
The dominant cost per global round remains local training at each client (Phase 0), which is identical to standard federated learning: $\mathcal{O}(E \cdot |B_i| \cdot C)$, where $E$ is the number of local epochs, $|B_i|$ is the batch size, and $C$ denotes the cost of a forward-backward pass through the CNN.

The additional defense mechanisms introduce the following overhead:
\begin{itemize}
    \item DDP Gaussian perturbation (Phase 0): $\mathcal{O}(d)$ per client, where $d$ is the model dimension.
    \item PLR extraction (Phase 1): $\mathcal{O}(d_p)$ per model, with $d_p \ll d$ being the penultimate layer dimension.
    \item AutoEncoder encoding for $n$ local models and one global model: $\mathcal{O}(n \cdot d_p \cdot h)$, where $h$ is the hidden size of the lightweight AE.
    \item CKA computation (linear kernel) for $n$ clients: $\mathcal{O}(n \cdot d_{lsr}^2)$, where $d_{lsr}$ is the compressed LSR dimension.
    \item K-Means clustering on $n$ CKA scores with $k=2$: $\mathcal{O}(n \cdot iter)$ (negligible, as $iter < 10$).
\end{itemize}

Overall per-round overhead of the latent-space defense and blockchain validation is $\mathcal{O}(n \cdot d_p \cdot h + n \cdot d_{lsr}^2)$, which is significantly smaller than local training cost. Smart-contract execution on the 6-node Hyperledger Fabric network adds only constant overhead per transaction due to the compact input (CKA scores and metadata).

\textbf{Communication Overhead:}
Each client uploads exactly one DDP-perturbed model $\tilde{w}_i^t$ per round (size $\approx d$), the same as vanilla FL. Additional messages consist of signatures, noise parameter $\sigma_i$, and hash pointers, contributing $\mathcal{O}(1)$ per client. The global model broadcast and blockchain consensus occur only within the small permissioned network (6 nodes), not among all $n$ clients. Thus, total communication complexity remains $\mathcal{O}(n \cdot d)$ per round, comparable to centralized FL but without single-server bottleneck and with validation performed locally within the trusted Fabric network.

\textbf{Storage Overhead:}
Raw model weights are stored off-chain via IPFS (distributed and content-addressable), while the blockchain ledger $\mathcal{L}$ only maintains compact metadata and hashes: $\mathcal{O}(1)$ storage per update. This results in linear growth of on-chain storage with the number of training rounds, independent of model size. Each edge node only needs to maintain its local model and the latest global model, keeping per-node storage modest and scalable for long-term industrial deployments.

In summary, PenTiDef introduces only marginal additional overhead compared to standard FL while delivering substantial gains in privacy, robustness, and decentralization. The computational, communication, and storage costs are acceptable for typical IIoT edge gateways and industrial servers, making the framework suitable for practical deployment in heterogeneous Cyber-Physical Systems where both security and operational efficiency are required.

\section{Research Questions and Experimental Design} \label{sec:researchquestion}
\subsection{Research Questions}
In this section, we address the following research questions (RQ):

\begin{itemize}
    \item \textbf{RQ1}: To what extent does the application of Distributed Differential Privacy (DDP) affect the model’s training performance?
    
    \item \textbf{RQ2}: How does the PenTiDef framework perform under non-IID data distributions, and how does this performance compare to IID settings?
    
    \item \textbf{RQ3}: How does PenTiDef compare with existing state-of-the-art defense mechanisms in terms of detection accuracy and robustness?
    
    \item \textbf{RQ4}: How effective is PenTiDef in detecting both untargeted and targeted poisoning attacks under different adversarial conditions?
    
    \item \textbf{RQ5}: Does PenTiDef achieve cost-efficiency in terms of computational overhead and training time compared to baseline approaches?
    
    \item \textbf{RQ6}: Is PenTiDef robust across different DL architectures, and does it maintain consistent performance?
    
    \item \textbf{RQ7}: What is the transaction processing capability of the blockchain coordination layer under different throughput and latency conditions?
\end{itemize}


\subsection{Experimental Scenarios}
To address the research questions above, we design four experimental scenarios as follows:

\subsubsection{Scenario 1: Impact of DDP on Model Performance (RQ1)}

This experiment assesses the effect of applying DDP noise to local models. Specifically, we evaluate PenTiDef with and without DDP under controlled noise levels ($\sigma \in [0, 0.2]$) to determine its impact on convergence and detection accuracy.

\subsubsection{Scenario 2: Comparative Evaluation under IID and non-IID Settings (RQ2, RQ3, RQ4, RQ5)}

This scenario compares PenTiDef with two representative defense mechanisms—FLARE~\cite{wang2022flare} and FedCC~\cite{fedcc}—under both IID and non-IID data distributions. We simulate poisoning attacks (as defined in Section~\ref{attack_strategy}) with varying adversary ratios (10\%, 20\%, 40\%) and evaluate the models using multiple performance metrics. Training time is also recorded to assess cost-efficiency.

\subsubsection{Scenario 3: Cross-Architecture Robustness Evaluation (RQ5, RQ6)}

To examine model generalizability, we repeat Scenario 2 using different DL architectures—CNN and SqueezeNet. This allows us to evaluate PenTiDef’s stability across heterogeneous model backbones.

\subsubsection{Scenario 4: Blockchain Layer Performance Benchmarking (RQ7)}

This scenario evaluates the transaction processing capacity of the Hyperledger-based blockchain layer used for decentralized coordination. We measure latency, throughput, and success rate under increasing transaction loads (5, 20, and 50 TPS) by executing smart contract operations such as model submission, querying, and verification.

\section{Implementation and Evaluation metrics} \label{sec:implementation}
\label{sec:metrics}

\begin{figure*}[!htp]
    \centering
    \includegraphics[width=0.95\textwidth]{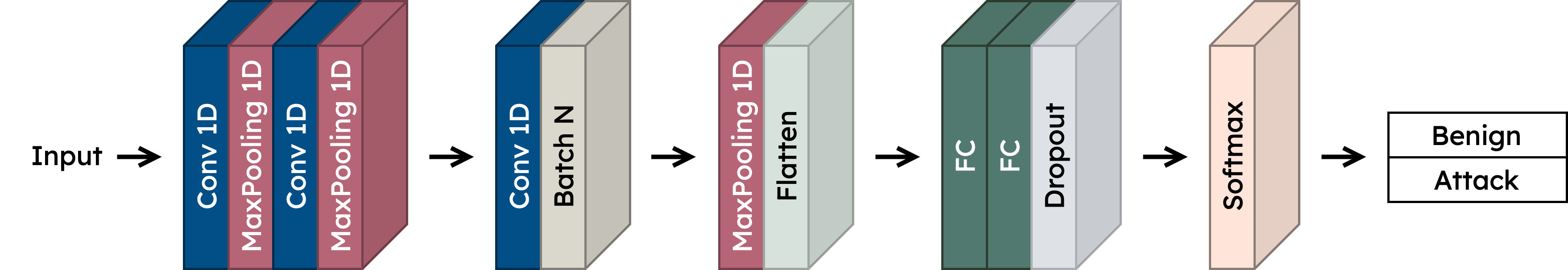}
    \caption{Details of layers in CNN}
    \label{fig:CNN}
\end{figure*}

\subsection{Environmental Setup}
In this study, our experimental environment is set up on a computer running Windows OS, equipped with an AMD Ryzen 7 6800HS CPU, 16GB DDR RAM.

In our implementation, the blockchain coordination layer is realized using Hyperledger Fabric, a permissioned blockchain platform that supports modular consensus and fine-grained access control, enabling secure, auditable, and efficient orchestration of decentralized FL processes. For off-chain model parameter storage, PenTiDef integrates the InterPlanetary File System (IPFS) to efficiently manage and retrieve local model updates based on their corresponding hash references maintained on the blockchain ledger.

\subsection{Blockchain Architecture and Performance Evaluation Framework}

In a Hyperledger Fabric network, multiple organizations collaborate to develop and maintain the blockchain system. Each organization typically operates its own certificate authority (CA), which issues identity certificates to both users and system components. Peer nodes, authenticated through their respective organizations’ CAs, maintain copies of the ledger—a distributed and tamper-resistant record of all transactions. These nodes are responsible for validating transactions, executing smart contracts (chaincode), and updating their local ledgers accordingly. In addition to peer nodes, the network incorporates an ordering service that sequences transactions chronologically before committing them to the ledger, ensuring consistency across the system.

Applications interact with the network through a client software development kit, sending transaction requests to peer nodes and invoking smart contracts to perform operations or query the ledger. Transaction proposals are signed using the user’s identity certificate, endorsed by selected peer nodes, and then submitted to the ordering service. The ordering service packages proposals into blocks, which are distributed to peer nodes for validation and ledger updates.

In our study, the network is composed of 3 organizations, each one contains 2 peer nodes. Moreover, to evaluate the network’s performance, we employed Hyperledger Caliper, a benchmarking framework developed by the Linux Foundation that provides standardized and trustworthy performance metrics fully compatible with Hyperledger Fabric. Caliper was used to measure key indicators such as transaction throughput (Transactions Per Second—TPS), transaction latency, and system resource utilization (CPU and memory consumption), enabling a comprehensive assessment of the system’s deployment efficiency and transaction performance within the Hyperledger Fabric environment.

\subsection{Neural Networks for IDS model}
We use a Convolutional Neural Network (CNN) to carry out the experiment. The model comprises one input layer, 11 hidden layers, and one output layer, as shown in \textbf{Figure \ref{fig:CNN}}. The input layer processes raw data for feature extraction, while the output layer generates classification results. Our model employs the Adam optimizer and the ReLU activation function. Training involves a batch size of 1024 and runs for 5 epochs, with the FL simulation incorporating 20 clients.

\subsection{Dataset}
\begin{figure*}
    \centering
    \begin{subfigure}{0.9\textwidth}
        \centering
        \includegraphics[width=\linewidth]{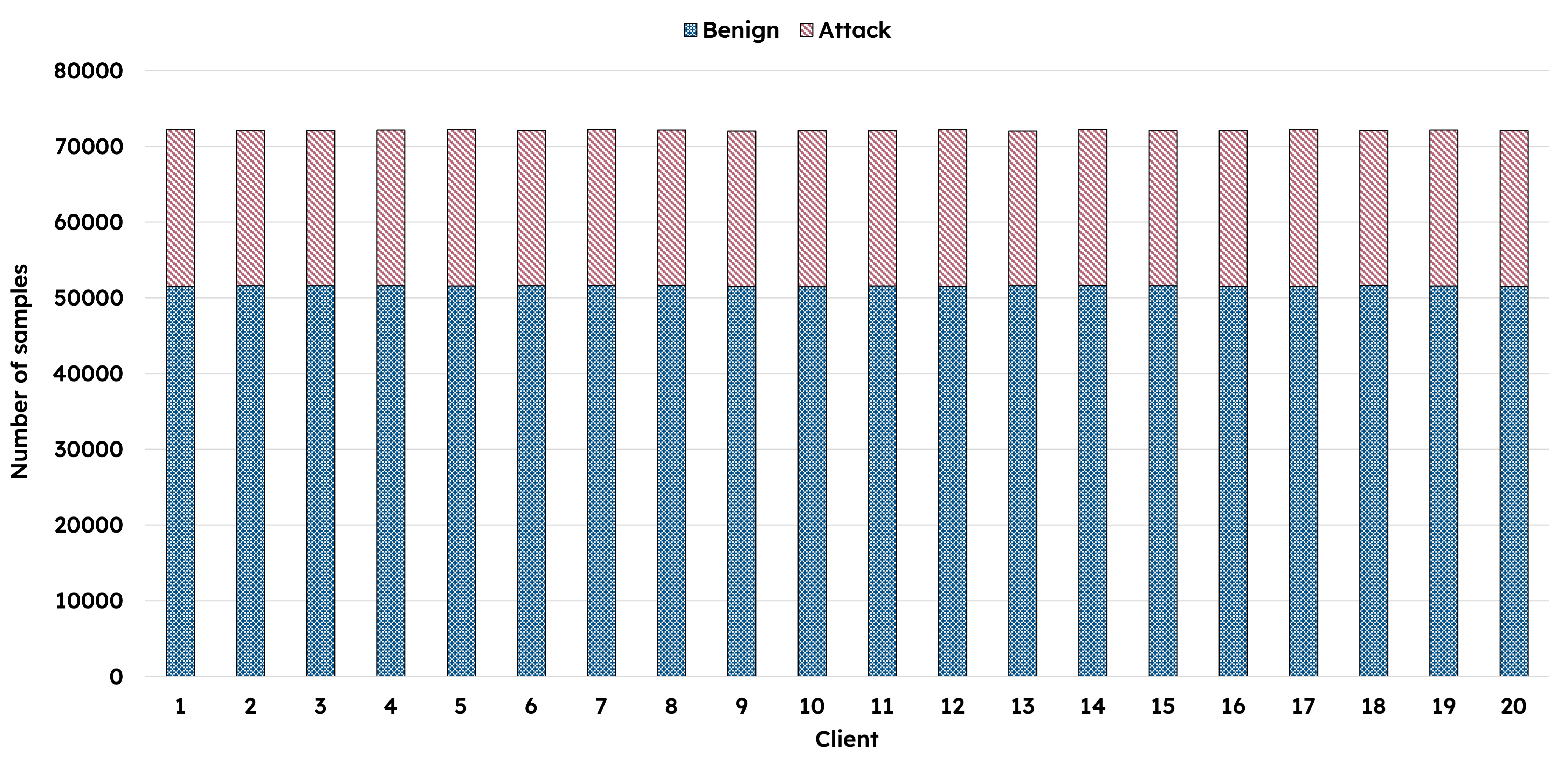}
        \caption{IID}
        \label{fig:cic:iid}
    \end{subfigure}
    \hfill
    \begin{subfigure}{0.9\textwidth}
        \centering
        \includegraphics[width=\linewidth]{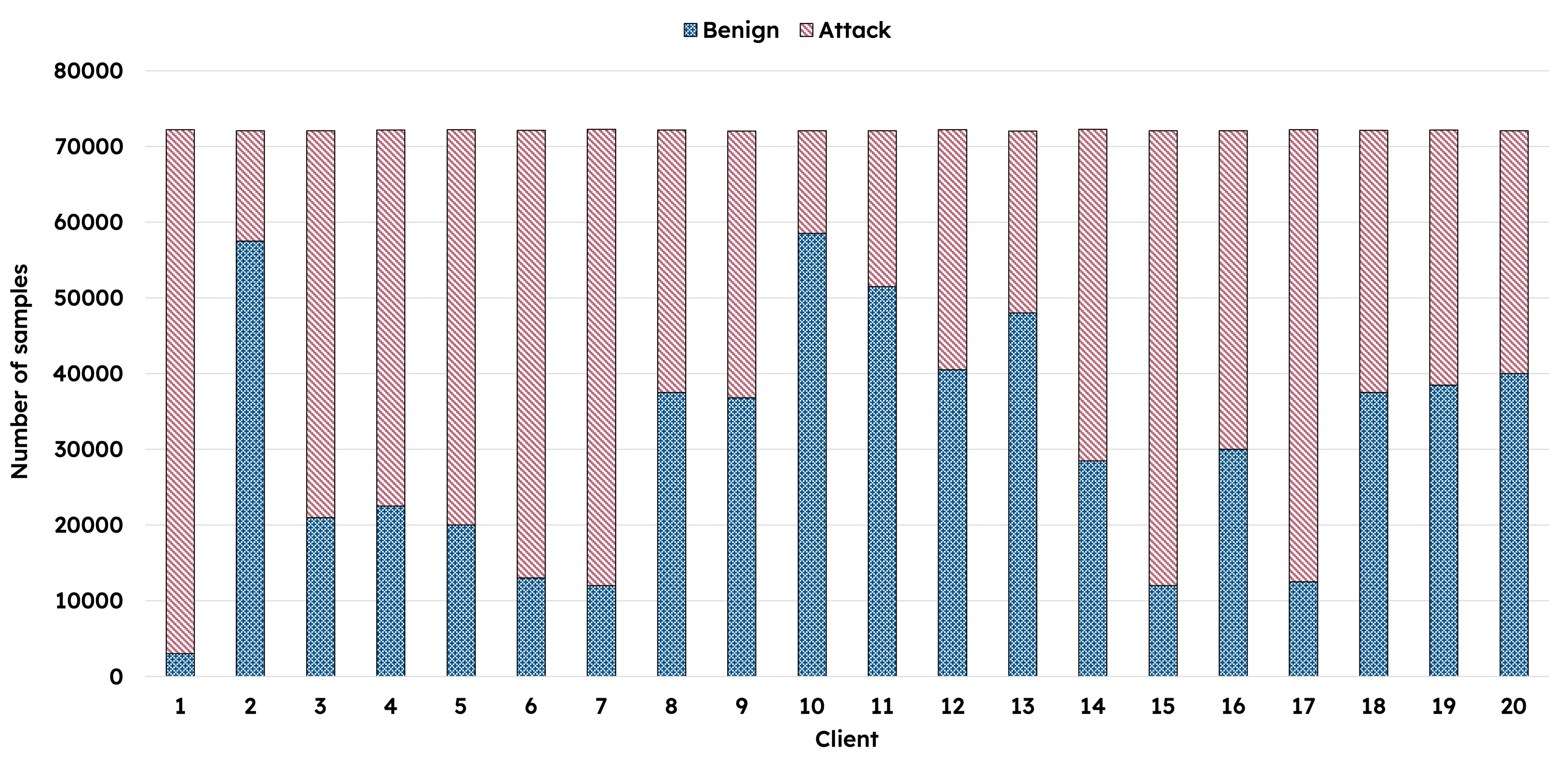}
        \caption{non-IID}
        \label{fig:cic:non-iid}
    \end{subfigure}
    \caption{Data distribution of the Edge-IIoTSet dataset under IID and non-IID scenarios}
    \label{fig:cic}
\end{figure*}

\begin{figure*}
    \centering
    \begin{subfigure}{0.9\textwidth}
        \centering
        \includegraphics[width=\linewidth]{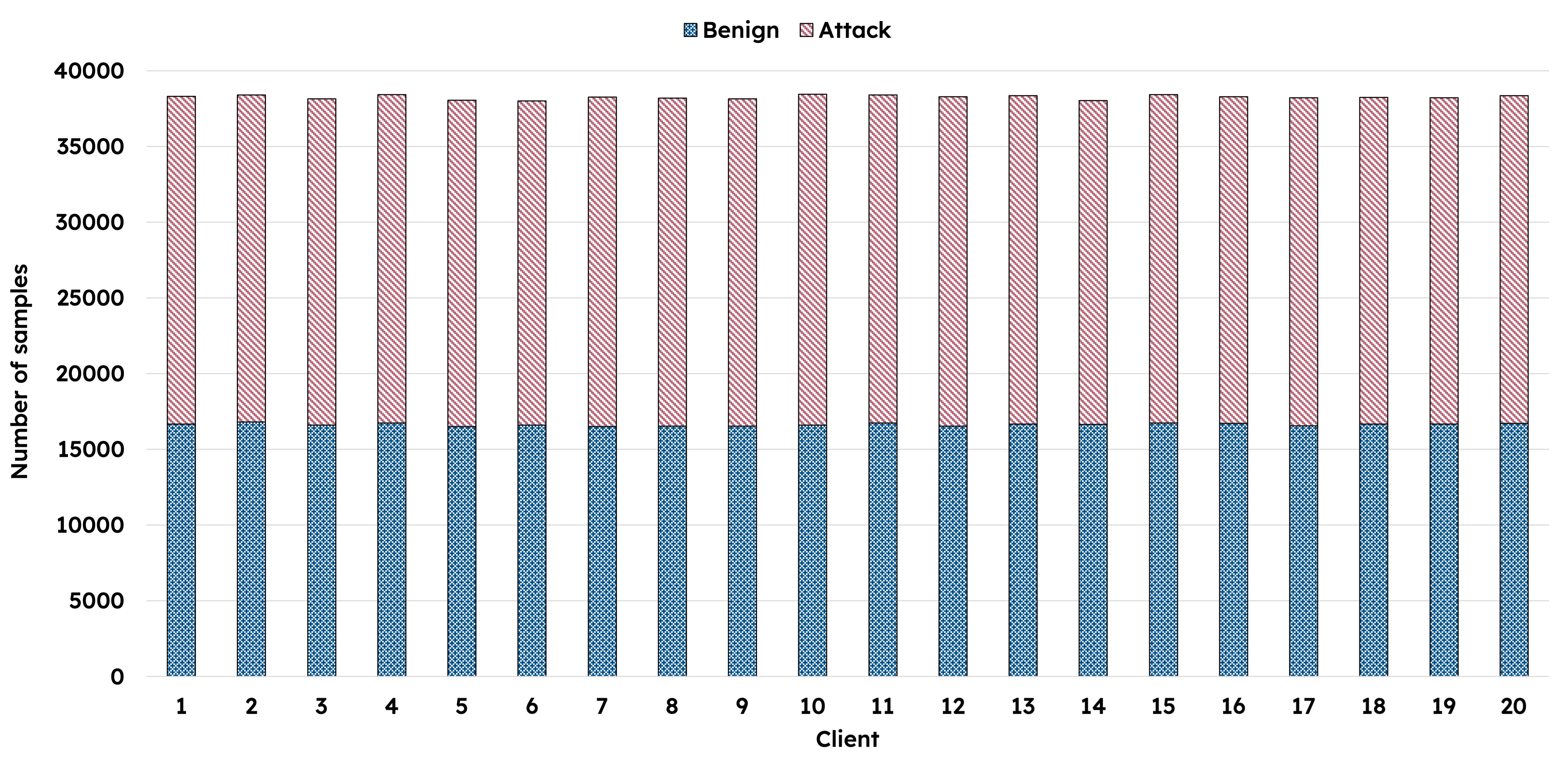}
        \caption{IID}
        \label{fig:edge:iid}
    \end{subfigure}
    \hfill
    \begin{subfigure}{0.9\textwidth}
        \centering
        \includegraphics[width=\linewidth]{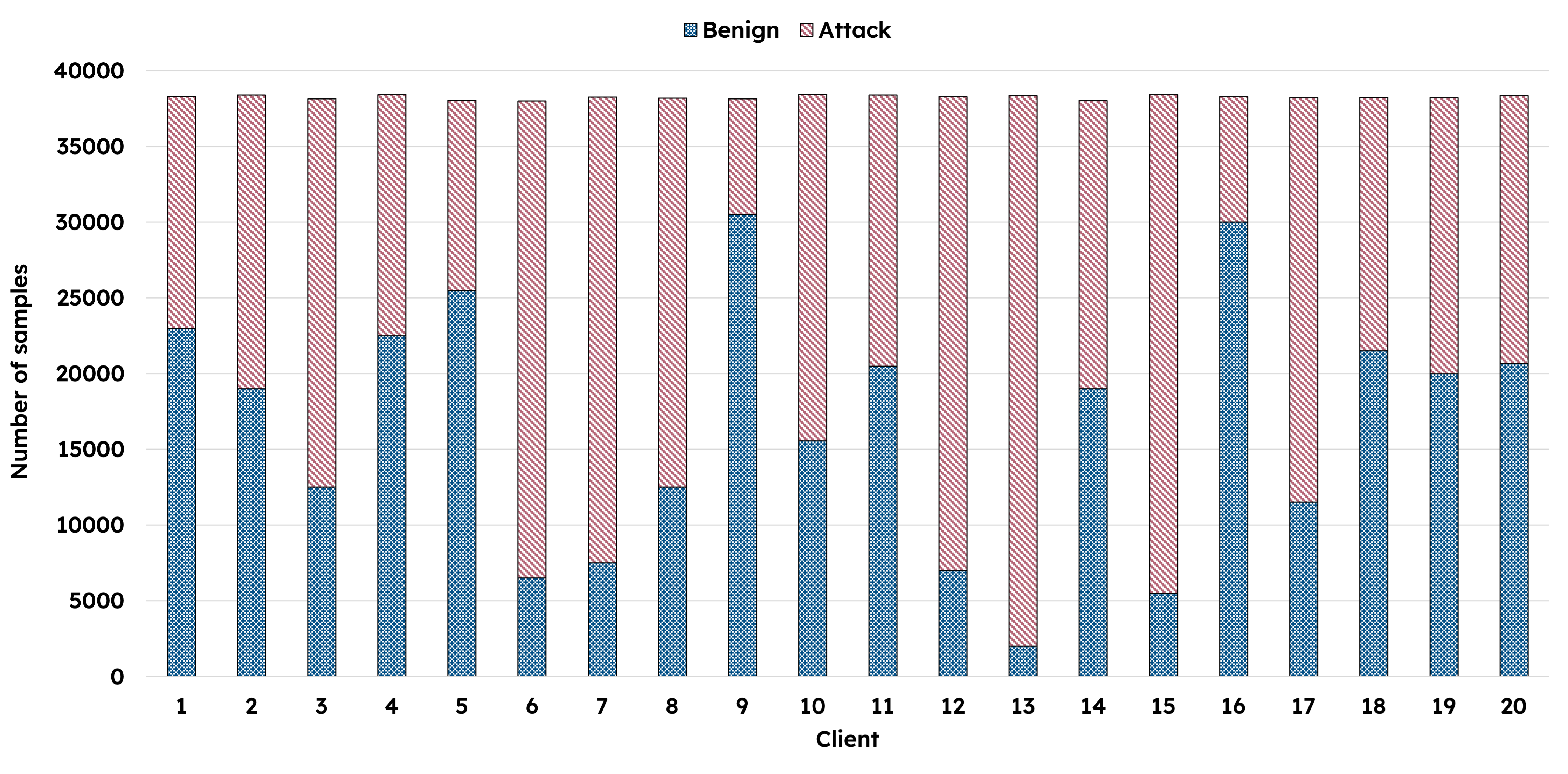}
        \caption{non-IID}
        \label{fig:edge:non-iid}
    \end{subfigure}
    \caption{Data distribution of the CIC-IDS2018 dataset under IID and non-IID scenarios}
    \label{fig:edge}
\end{figure*}

Our benchmarking scenarios utilize two datasets: Edge-IIoTSet \cite{mbc1-1h68-22} and CIC-IDS2018 \cite{sharafaldin2018toward}. After preprocessing, the Edge-IIoTSet dataset consists of 95 features and one label column, with a labeling ratio of roughly 71.4\% benign data and 28.6\% attack data. In the case of the CIC-IDS2018 dataset, there are 71 features and a label column after preprocessing, with approximately 42.6\% benign data and 57.4\% attack data. We assign 30\% of each dataset for testing to ensure an unbiased evaluation of the performance of the model, while the remaining 70\% was evenly distributed among the collaborating machines.
In addition, we have divided the datasets into IID and non-IID subsets to thoroughly evaluate the performance and robustness of the PenTiDef model under different data distributions. In doing so, our objective is to demonstrate the model’s capability to maintain high accuracy and stability, regardless of the underlying data distribution, further validating its applicability and reliability in diverse FL applications. The details of the data distribution for the two datasets on iid and non-iid condition are shown in \textbf{Figure \ref{fig:cic}} and \textbf{Figure \ref{fig:edge}}.

\subsection{Evaluation Metrics}

To evaluate the effectiveness of the proposed PenTiDef framework, we adopt four standard metrics commonly used in binary classification tasks: Accuracy, Precision, Recall, and F1-Score.

\textit{Accuracy} measures the overall proportion of correctly classified samples.  
\textit{Precision} reflects the proportion of correctly predicted positive instances among all predicted positives.  
\textit{Recall} quantifies the model's ability to identify actual positive instances.  
\textit{F1-Score} represents the harmonic mean of Precision and Recall, offering a balanced measure in cases of class imbalance.

These metrics are derived from the confusion matrix composed of true positives, true negatives, false positives, and false negatives. They provide a comprehensive view of model performance in distinguishing between benign and malicious traffic. In addition, we employ the \textit{Centered Kernel Alignment (CKA) Score} to measure the similarity between latent representations of local and global models. This metric is particularly useful for evaluating the consistency of feature space across participants and identifying anomalies introduced by poisoned updates. Based on this measure, we can distinguish and eliminate poisoned models from the update process. A higher CKA score indicates that the local model is closer to the global model. 

\section{Experimental Results and Analysis} \label{sec:results}
\subsection{Result 1 (Answer to Scenario 1): PenTiDef maintains high performance under DDP with minimal accuracy degradation}
\begin{figure*}[!htb]
    \centering
    \includegraphics[width=13cm]{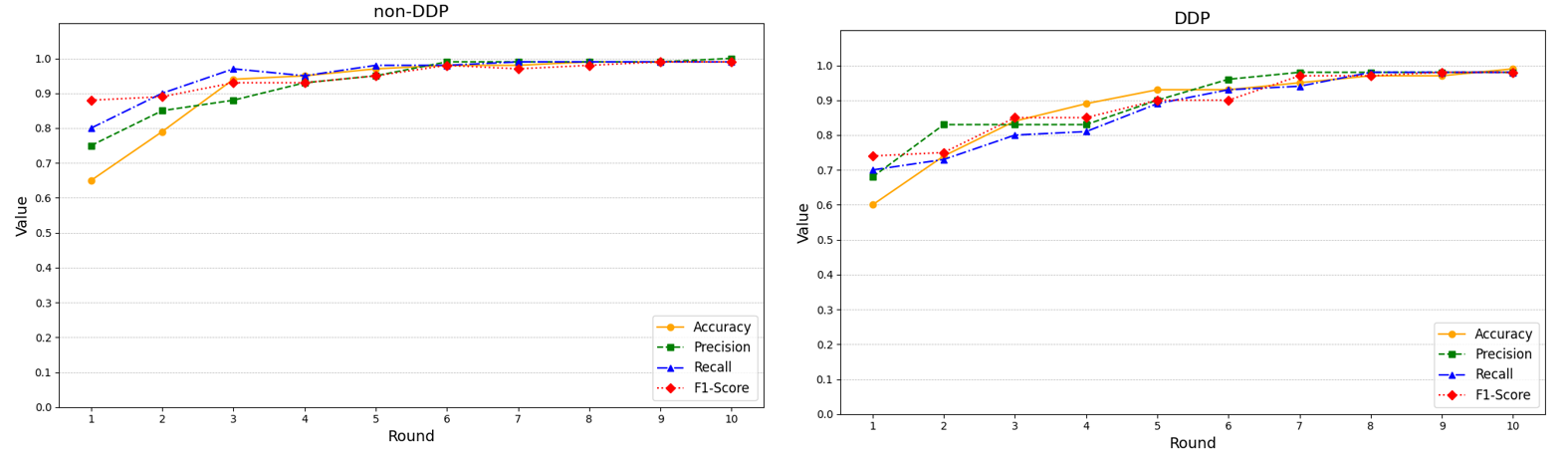}
    \caption{Comparison of training results with and without applying DDP}
    \label{fig:ddp_comparision}
\end{figure*}


\begin{table*}[!htp]
\centering
\caption{Performance comparison of defense methods on IID datasets for detecting untargeted attacks}
\begin{adjustbox}{width=\textwidth}
\begin{tabular}{cccccccccccccc}
\toprule
 &  & \multicolumn{4}{c}{\textbf{Adversaries 10\%}} & \multicolumn{4}{c}{\textbf{Adversaries 20\%}} & \multicolumn{4}{c}{\textbf{Adversaries 40\%}} \\ \toprule
Dataset & Attack & No-Def & Flare & FedCC & \textbf{PenTiDef}  & No-Def & Flare & FedCC & \textbf{PenTiDef}  & No-Def & Flare & FedCC & \textbf{PenTiDef}  \\ \toprule
\multirow{4}{*}{CIC-IDS2018} & LF & 0.77 & 0.96 & 0.96 & \textbf{{0.97}} & 0.75 & 0.93 & 0.95 & \textbf{{0.96}} & 0.71 & 0.92 & 0.95 & \textbf{{0.96}} \\ \cmidrule{2-14}
 & WS & 0.76 & 0.95 & 0.95 & \textbf{{0.96}} & 0.76 & \textbf{{0.97}} & 0.95 & 0.96 & 0.68 & 0.92 & 0.96 & \textbf{{0.96}} \\ \cmidrule{2-14}
 & Un-Krum & 0.73 & 0.95 & \textbf{{0.97}} & 0.96 & 0.72 & 0.95 & 0.96 & \textbf{{0.97}} & 0.64 & \textbf{{0.96}} & 0.88 & 0.93 \\ \cmidrule{2-14}
 & Un-Med & 0.73 & 0.96 & 0.96 & \textbf{{0.98}} & 0.65 & 0.95 & 0.93 & \textbf{{0.98}} & 0.67 & 0.91 & 0.92 & \textbf{{0.97}} \\ \hline
\multirow{4}{*}{Edge-IIoTset} & LF & 0.65 & \textbf{0.99} & 0.98 & 0.98  & 0.63 & 0.92 & 0.96 & \textbf{{0.97}} & 0.57 & 0.88 & 0.88 & \textbf{{0.92}} \\ \cmidrule{2-14}
 & WS & 0.64 & 0.95 & \textbf{{0.96}} & 0.95 & 0.63 & 0.95 & \textbf{{0.97}} & 0.95 & 0.58 & 0.91 & 0.88 & \textbf{{0.95}} \\ \cmidrule{2-14}
 & Un-Krum & 0.67 & 0.94 & 0.94 & \textbf{{0.95}} & 0.67 & \textbf{{0.98}} & 0.91 & 0.96 & 0.58 & 0.91 & 0.87 & \textbf{{0.93}} \\ \cmidrule{2-14}
 & Un-Med & 0.73 & 0.94 & 0.96 & \textbf{{0.97}} & 0.64 & 0.88 & 0.93 & \textbf{{0.99}} & 0.56 & 0.88 & 0.88 & \textbf{{0.95}} \\ \bottomrule
\end{tabular}
\end{adjustbox}

\label{tab:untargeted_iid}
\end{table*}

\begin{table*}[!htp]
\centering
\caption{Performance comparison of defense methods on IID datasets for detecting targeted attacks}
\begin{adjustbox}{width=\textwidth}
\begin{tabular}{cccccccccccccc}
\toprule
 &  & \multicolumn{4}{c}{\textbf{Adversaries 10\%}} & \multicolumn{4}{c}{\textbf{Adversaries 20\%}} & \multicolumn{4}{c}{\textbf{Adversaries 40\%}} \\ \toprule
Dataset & Attack & No-Def & Flare & FedCC & \textbf{PenTiDef}  & No-Def & Flare & FedCC & \textbf{PenTiDef}  & No-Def & Flare & FedCC & \textbf{PenTiDef}  \\ \toprule
\multirow{4}{*}{CIC-IDS2018} & GAN-SL & 0.68 & 0.96 & 0.95 & \textbf{0.97} & 0.67 & 0.93 & 0.95 & \textbf{0.97} & 0.57 & 0.90 & 0.93 & \textbf{0.94}  \\ \cmidrule{2-14}
 & GAN-ML & 0.73 & \textbf{0.97} & 0.96 & 0.96 & 0.73 & 0.95 & \textbf{0.96} & \textbf{0.96} & 0.63 & 0.93 & 0.96 & \textbf{0.97} \\ \cmidrule{2-14}
 & GAN-Con & 0.74 & 0.95 & 0.96 & \textbf{0.97} & 0.74 & 0.95 & 0.96 & \textbf{0.97} & 0.56 & 0.88 & 0.93 & \textbf{0.95} \\ \cmidrule{2-14}
 & BD & 0.69 & 0.96 & 0.96 & \textbf{0.97} & 0.69 & 0.96 & 0.96 & \textbf{0.97} & 0.59 & 0.94 & 0.92 & \textbf{0.94}\\ \hline
\multirow{4}{*}{Edge-IIoTset} & GAN-SL & 0.58 & 0.93 & 0.92 & 0.92 & 0.58 & \textbf{0.93} & 0.92 & \textbf{0.93} & 0.58 & 0.93 & 0.93 & \textbf{0.94} \\ \cmidrule{2-14}
 & GAN-ML & 0.63 & 0.93 & \textbf{0.94} & 0.93 & 0.63 & 0.93 & \textbf{0.95} & 0.93 & 0.63 & 0.92 & 0.93 & \textbf{0.94}\\ \cmidrule{2-14}
 & GAN-Con & 0.60 & 0.92 & 0.92 & \textbf{0.93} & 0.60 & \textbf{0.96} & 0.90 & 0.95 & 0.60 & 0.93 & 0.92 & \textbf{0.94} \\ \cmidrule{2-14}
 & BD & 0.59 & 0.92 & 0.93 & \textbf{0.93} & 0.59 & 0.88 & 0.91 & \textbf{0.93} & 0.59 & 0.94 & 0.94 & \textbf{0.95} \\ \bottomrule
\end{tabular}
\end{adjustbox}
\label{tab:Targeted_iid}
\end{table*}

\begin{table*}[!htp]
\centering
\caption{Performance comparison of defense methods on non-IID datasets for detecting untargeted attacks}
\begin{adjustbox}{width=\textwidth}
\begin{tabular}{cccccccccccccc}
\toprule
 &  & \multicolumn{4}{c}{\textbf{Adversaries 10\%}} & \multicolumn{4}{c}{\textbf{Adversaries 20\%}} & \multicolumn{4}{c}{\textbf{Adversaries 40\%}} \\ \toprule
Dataset & Attack & No-Def & Flare & FedCC & \textbf{PenTiDef}  & No-Def & Flare & FedCC & \textbf{PenTiDef}  & No-Def & Flare & FedCC & \textbf{PenTiDef}  \\ \toprule
\multirow{4}{*}{CIC-IDS2018} & LF & 0.36 & 0.92 & 0.93 & \textbf{{0.95}} & 0.46 & 0.91 & 0.93 & \textbf{{0.94}} & 0.41 & 0.90 & 0.92 & \textbf{{0.93}} \\ \cmidrule{2-14}
 & WS & 0.42 & 0.93 & 0.93 & \textbf{{0.94}} & 0.47 & \textbf{{0.93}} & 0.91 & 0.92 & 0.41 & 0.90 & 0.92 & \textbf{{0.93}} \\ \cmidrule{2-14}
 & Un-Krum & 0.40 & 0.90 & 0.93 & \textbf{{0.94}} & 0.44 & 0.92 & 0.92 & \textbf{{0.93}} & 0.40 & \textbf{{0.92}} & 0.84 & 0.89 \\ \cmidrule{2-14}
 & Un-Med & 0.43 & 0.91 & 0.91 & \textbf{{0.94}} & 0.44 & 0.92 & 0.92 & \textbf{{0.93}} & 0.40 & 0.87 & 0.88 & \textbf{{0.95}} \\ \midrule
\multirow{4}{*}{Edge-IIoTset} & LF & 0.37 & \textbf{0.91} & 0.90 & 0.90 & 0.37 & 0.92 & 0.93 & \textbf{{0.94}} & 0.37 & 0.84 & 0.85 & \textbf{{0.88}} \\ \cmidrule{2-14}
 & WS & 0.44 & 0.89 & \textbf{{0.91}} & 0.91 & 0.41 & 0.91 & \textbf{{0.93}} & 0.92 & 0.44 & 0.87 & 0.86 & \textbf{{0.91}} \\ \cmidrule{2-14}
 & Un-Krum & 0.46 & 0.90 & 0.90 & \textbf{{0.91}} & 0.43 & \textbf{{0.94}} & 0.89 & 0.92 & 0.44 & 0.87 & 0.83 & \textbf{{0.89}} \\ \cmidrule{2-14}
 & Un-Med & 0.51 & 0.91 & 0.89 & \textbf{{0.94}} & 0.40 & 0.84 & 0.89 & \textbf{{0.95}} & 0.42 & 0.84 & 0.84 & \textbf{{0.91}} \\ \bottomrule
\end{tabular}
\end{adjustbox}
\label{tab:untargeted_non_iid}
\end{table*}

The performance of the PenTiDef model under two conditions, without applying DDP (non-DDP) and with applying DDP, is shown in \textbf{Figure \ref{fig:ddp_comparision}}, demonstrating a distinct difference in the initial stage. Initially, when applying DDP, the training performance is lower than non-DDP due to DDP adding noise to the model to protect privacy, resulting in a slower convergence rate.

In the early training rounds, the PenTiDef model with DDP shows slower growth and lower metrics compared to non-DDP. However, in the subsequent rounds, the DDP model gradually converges and reaches stability, though still approximately 0.01 lower than non-DDP across all criteria.

These results demonstrate that PenTiDef maintains high and stable performance even when applying DDP, reflecting a balance between security and performance. PenTiDef is capable of adapting and operating efficiently in decentralized FL environments, ensuring security while maintaining high accuracy in detecting network threats.


\begin{figure*}[!b]
    \centering
    \includegraphics[width=0.9\linewidth]{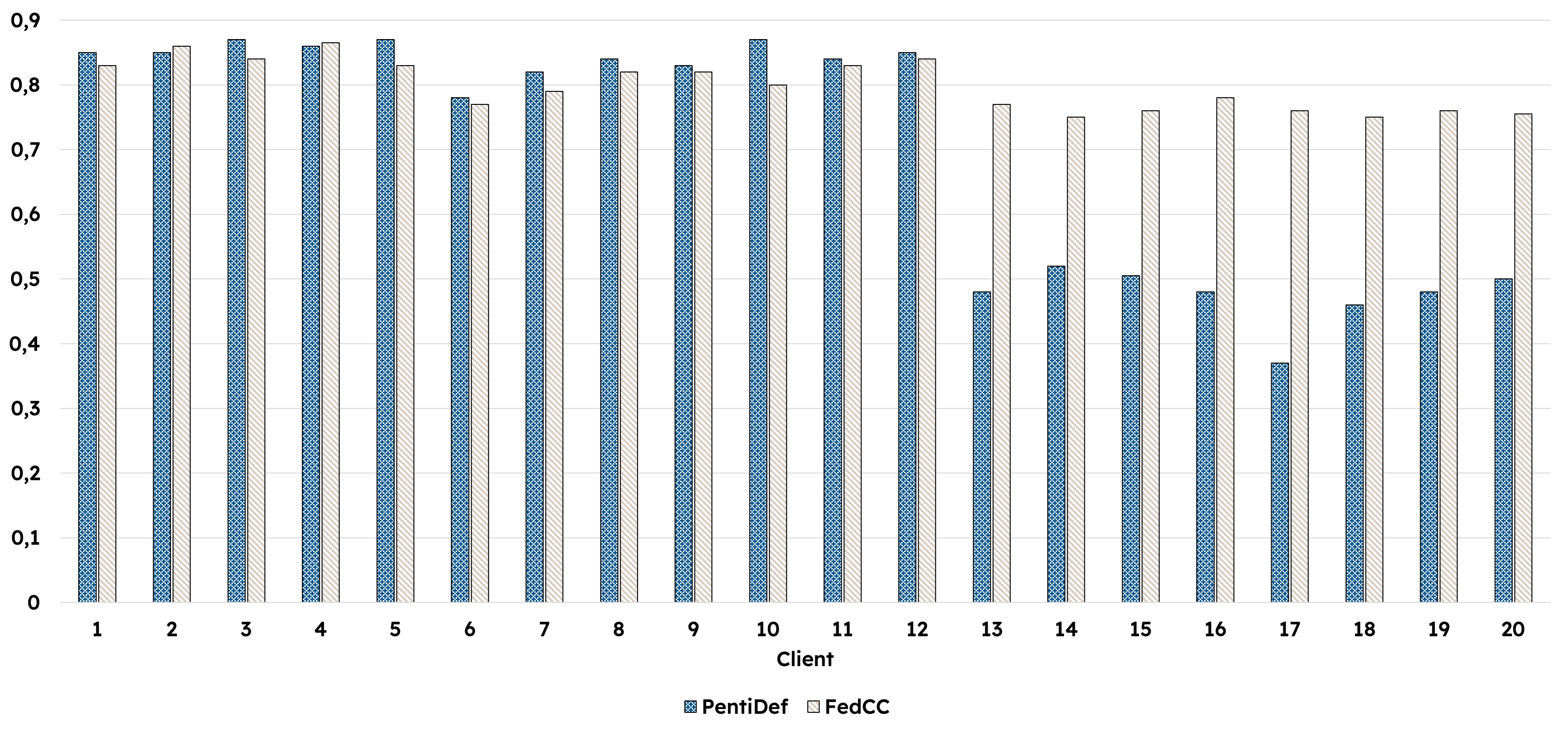}
    \caption{Similarity comparison between the latent space of the global model and each local model's latent space using CKA scores in PenTiDef and FedCC.}
    \label{fig:cka_untar_iid}
\end{figure*}

\subsection{Result 2 (Answer to Scenario 2): PenTiDef outperforms state-of-the-art defenses across IID and non-IID data under multiple poisoning attacks}

\textbf{Training on IID Datasets}
The results in \textbf{Table \ref{tab:untargeted_iid}} demonstrate that PenTiDef outperforms other models in most cases for untargeted attacks. Specifically, PenTiDef consistently achieves the highest or near-highest values in evaluation criteria such as Accuracy, Precision, Recall, and F1-Score. This is evident across all levels of adversarial presence, showing that PenTiDef maintains stable and superior performance even as the adversary rate increases.
PenTiDef's performance remains robust when facing various types of attacks, especially in cases like Untargeted-Med and Un-Krum, where it frequently achieves the highest performance. This highlights the effectiveness of PenTiDef in detecting and defending against untargeted poisoning attacks.

\textbf{Table \ref{tab:Targeted_iid}} lists the parameters of the defense methods against targeted attacks. Overall, all three defense methods show a comparison to detecting untargeted attacks. This can be explained by the fact that targeted attacks are harder to detect. Targeted attacks are designed specifically for each target, utilizing evasion techniques, exploiting zero-day vulnerabilities, and employing APTs. Attackers often patiently gather intelligence, maintain long-term persistence, and leverage insider threats, making them significantly more challenging to identify. In contrast, untargeted attacks are more widespread, relying on common tools and techniques, which makes them easier to detect. Targeted attacks often have sophisticated characteristics and specific goals, such as skewing predictions for a particular type of input without significantly altering the model's overall performance. These small and subtle changes require defense methods to be highly sensitive. Nonetheless, our model still achieves the highest performance in most of these attacks.


\begin{table*}[!htb]
\centering
\caption{Performance comparison of defense methods on non-IID datasets for detecting targeted attacks}
\begin{adjustbox}{width=\textwidth}
\begin{tabular}{cccccccccccccc}
\toprule
& & \multicolumn{4}{c}{\textbf{Adversaries 10\%}} & \multicolumn{4}{c}{\textbf{Adversaries 20\%}} & \multicolumn{4}{c}{\textbf{Adversaries 40\%}} \\ \toprule
Dataset & Attack & No-Def & Flare & FedCC & \textbf{PenTiDef}  & No-Def & Flare & FedCC & \textbf{PenTiDef}  & No-Def & Flare & FedCC & \textbf{PenTiDef}  \\ \toprule
\multirow{4}{*}{CIC-IDS2018} & GAN-SL & 0.55 & 0.91 & 0.91 & \textbf{{0.92}} & 0.45 & 0.91 & 0.91 & \textbf{{0.92}} & 0.52 & 0.88 & 0.88 & \textbf{0.89} \\ \cmidrule{2-14}
 & GAN-ML & 0.53 & 0.92 & \textbf{{0.94}} & 0.92 & 0.45 & 0.91 & \textbf{{0.93}} & 0.92 & 0.50 & 0.89 & \textbf{0.91} & 0.89 \\ \cmidrule{2-14}
 & GAN-Con & 0.53 & 0.94 & 0.94 & \textbf{{0.95}} & 0.46 & \textbf{{0.93}} & \textbf{{0.93}} & 0.50 & 0.91 & 0.91 & \textbf{0.92} & \textbf{0.92} \\ \cmidrule{2-14}
 & BD & 0.54 & 0.93 & 0.93 & \textbf{{0.95}} & 0.46 & 0.93 & 0.92 & \textbf{{0.93}} & 0.51 & 0.90 & 0.90 & \textbf{0.92} \\ \midrule
\multirow{4}{*}{Edge-IIoTset} & GAN-SL & 0.55 & \textbf{{0.91}} & 0.89 & 0.89 & 0.46 & 0.91 & 0.89 & \textbf{{0.90}} & 0.52 & 0.88 & 0.86 & \textbf{0.86} \\ \cmidrule{2-14}
 & GAN-ML & 0.56 & 0.91 & \textbf{{0.93}} & 0.91 & 0.46 & 0.91 & \textbf{{0.93}} & 0.90 & 0.53 & 0.88 & \textbf{0.90} & 0.88 \\ \cmidrule{2-14}
 & GAN-Con & 0.56 & 0.92 & 0.92 & \textbf{{0.93}} & 0.44 & 0.92 & 0.92 & \textbf{{0.93}} & 0.53 & 0.89 & 0.89 & \textbf{0.90} \\ \cmidrule{2-14}
 & BD & 0.55 & 0.91 & 0.91 & \textbf{{0.95}} & 0.44 & 0.91 & 0.91 & \textbf{{0.94}} & 0.52 & 0.88 & 0.88 & \textbf{0.92} \\ \bottomrule
\end{tabular}
\end{adjustbox}
\label{tab:Targeted_non_iid}
\end{table*}


%
\begin{figure*}[!b]
    \centering
    \includegraphics[width=0.9\linewidth]{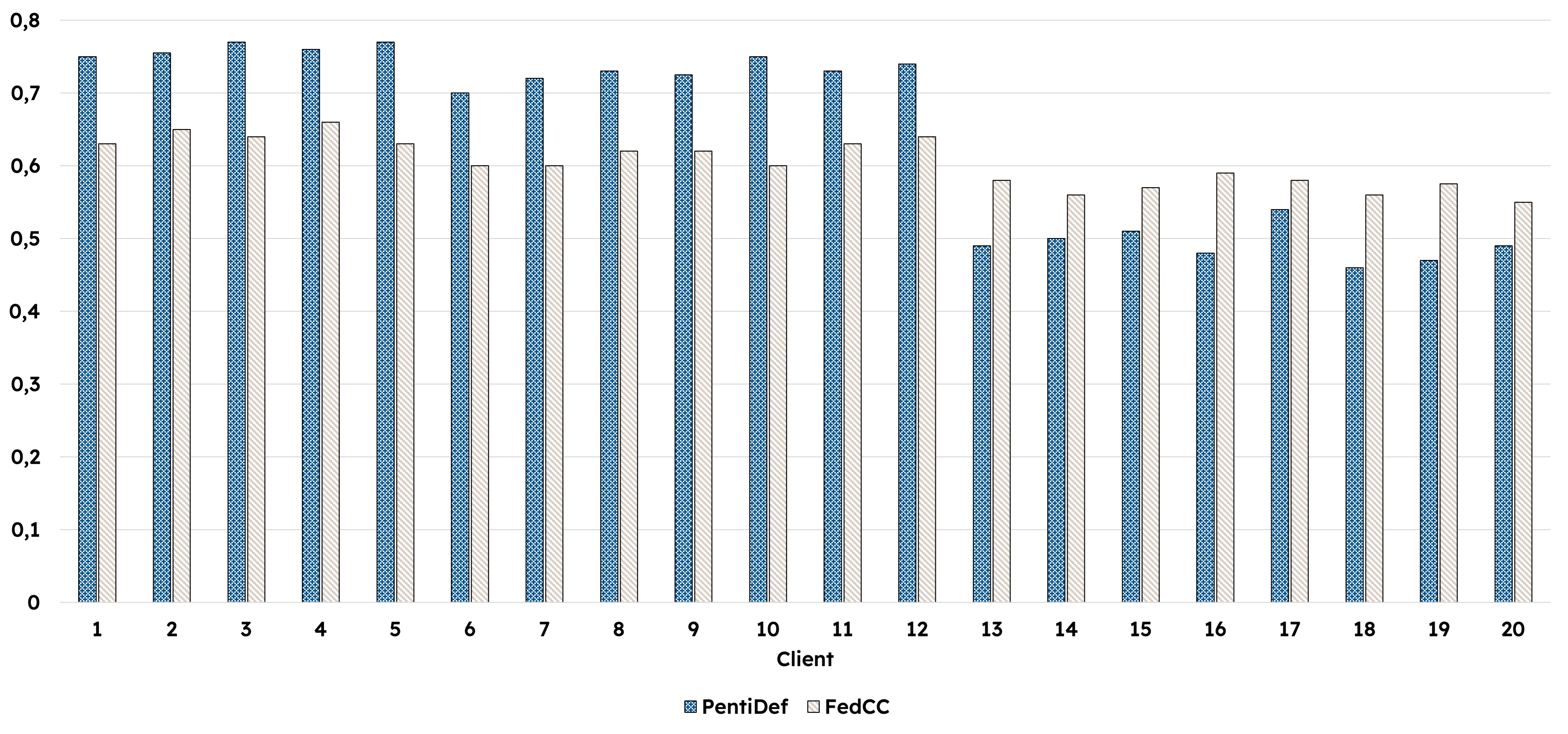}
    \caption{Compare the similarity between the latent space of the global model and each local model's latent space using CKA scores in PenTiDef and FedCC.}
    \label{fig:cka_non_iid}
\end{figure*}

To explain the stability in classifying benign and malicious models, as shown in \textbf{Figure \ref{fig:cka_untar_iid}}, we have compiled the CKA scores for each participant with an attacker ratio of 40\% (the last 8 collaborators) using the Untargeted-Krum attack method. These figures were recorded after the final training round. We compare the CKA scores of the two modules, PenTiDef and the module using a similar CKA scoring method, FedCC \cite{fedcc}.

It is easily noticeable that there is a significant difference between the CKA scores of benign and malicious LSRs compared to the LSR of the global model. For collaborators from clients 1 to 12 (assumed to be benign), PenTiDef maintains a high level of similarity with CKA scores mostly above 0.8, demonstrating its ability to maintain stability in identifying benign models. On the contrary, the CKA scores for collaborators from clients 13 to 20 (assumed to be malicious) show a significant decline, mostly below 0.6, with some even below 0.5. This indicates that PenTiDef can effectively distinguish between benign and malicious models, even with an attacker ratio of 40\%. Meanwhile, FedCC does not exhibit a clear distinction like PenTiDef, resulting in poorer classification performance. This explains why PenTiDef achieves higher performance in most scenarios compared to FedCC.

\textbf{Training on non-IID Dataset}
For the case of training on the non-IID heterogeneous dataset, the data distribution among the 20 collaborating models is divided as shown in \textbf{Figure \ref{fig:cic:non-iid} and Figure \ref{fig:edge:non-iid}}. Similar to the previous case, the following scenarios are simulated with attacker ratios of 10\%, 20\%, and 40\%. Using a non-IID dataset is crucial because, in real-world FL scenarios, data across different clients is inherently non-IID due to variations in user behavior, device environments, and data collection methods. This non-uniformity presents a greater challenge for defense mechanisms, as the differences in local data distributions can lead to model divergence and increased vulnerability to adversarial manipulations. Moreover, non-IID data can obscure attack patterns, making it harder for defenses to differentiate between natural variations in data and malicious manipulations. The reason for distributing the dataset in this manner is to clarify the differences between the defense methods in distinguishing malicious models when trained on a non-IID dataset. Regarding the results summarized in \textbf{Table \ref{tab:untargeted_non_iid}}, it can be observed that due to the influence of the heterogeneous data distribution, the performance results of all three defense modules experience a slight decline. However, our PenTiDef model still achieves the highest performance in most attack scenarios.

\begin{table}[!b]
\centering
\caption{Average training time on different datasets with different models}
\label{tab:training_time}
\begin{tabular}{ccc}
\toprule
\textbf{Dataset}    & \textbf{Model} & \textbf{Training Time} \\ \toprule
\multirow{3}{*}{CIC-IDS2018} & Flare & 3,907s $\sim$ 01h 06m \\ \cmidrule{2-3}
                          & FedCC & 2,460s $\sim$ 41m \\ \cmidrule{2-3}
                          & \textbf{PenTiDef} & \textbf{1,988s $\sim$ 33m} \\ \midrule
\multirow{3}{*}{Edge-IIoTset} & Flare & 4,395s $\sim$ 1h 15m \\ \cmidrule{2-3}
                          & FedCC & 2,825s $\sim$ 47m \\ \cmidrule{2-3}
                          & \textbf{PenTiDef} & \textbf{1,869s $\sim$ 31m} \\ \bottomrule
\end{tabular}

\end{table}


\begin{table*}[!t]
\centering
\caption{Performance Comparison of the PenTiDef module on the CIC-IDS2018 and Edge-IIoTset datasets with two DL models, CNN and SqueezeNet (untargeted attack)}
\begin{adjustbox}{width=\linewidth}
\begin{tabular}{ccccccccc}
\toprule
\multirow{2}{*}{\textbf{Dataset}} & \multirow{2}{*}{\textbf{Attack}} & \multicolumn{2}{c}{\textbf{10\% adv}} & \multicolumn{2}{c}{\textbf{20\% adv}} & \multicolumn{2}{c}{\textbf{40\% adv}} \\ \cmidrule{3-8} 
 &  & \textbf{CNN} & \textbf{SqueezeNet} & \textbf{CNN} & \textbf{SqueezeNet} & \textbf{CNN} & \textbf{SqueezeNet} \\ \toprule
\multicolumn{8}{c}{\textbf{IID}} \\ \toprule
\multirow{4}{*}{CIC-IDS2018} & LF & 0.97 & \textbf{0.98} & 0.97 & \textbf{0.98} & 0.94 & \textbf{0.96} \\ \cmidrule{2-8} 
 & WS & \textbf{0.96} & 0.95 & 0.96 & \textbf{0.97} & 0.96 & \textbf{0.97} \\ \cmidrule{2-8}
 & Un-Krum & 0.96 & \textbf{0.97} & 0.97 & \textbf{0.98} & 0.93 & \textbf{0.94} \\ \cmidrule{2-8}
 & Un-Med & \textbf{0.98} & 0.97 & \textbf{0.97} & 0.96 & 0.97 & \textbf{0.98} \\ \midrule
\multirow{4}{*}{Edge-IIoTset} & LF & \textbf{0.98} & 0.97 & 0.97 & \textbf{0.98} & 0.92 & \textbf{0.95} \\ \cmidrule{2-8} 
 & WS & \textbf{0.98} & 0.97 & 0.95 & \textbf{0.96} & 0.95 & \textbf{0.96} \\ \cmidrule{2-8}
 & Un-Krum & 0.94 & \textbf{0.95} & 0.93 & \textbf{0.94} & 0.89 & \textbf{0.90} \\ \cmidrule{2-8}
 & Un-Med & \textbf{0.98} & 0.97 & \textbf{0.99} & 0.98 & 0.95 & \textbf{0.97} \\ \bottomrule
\multicolumn{8}{c}{\textbf{non-IID}} \\ \toprule
\multirow{4}{*}{CIC-IDS2018} & LF & 0.95 & \textbf{0.97} & 0.94 & \textbf{0.95} & 0.93 & \textbf{0.95} \\ \cmidrule{2-8} 
 & WS & \textbf{0.94} & 0.93 & 0.92 & \textbf{0.94} & 0.93 & \textbf{0.94} \\ \cmidrule{2-8}
 & Un-Krum & \textbf{0.94} & 0.92 & 0.93 & \textbf{0.95} & \textbf{0.95} & 0.94 \\ \cmidrule{2-8}
 & Un-Med & \textbf{0.94} & 0.93 & \textbf{0.93} & 0.92 & \textbf{0.95} & 0.94 \\ \midrule
\multirow{4}{*}{Edge-IIoTset} & LF & \textbf{0.90} & 0.89 & \textbf{0.94} & 0.93 & 0.93 & \textbf{0.94} \\ \cmidrule{2-8} 
 & WS & 0.91 & \textbf{0.92} & \textbf{0.92} & 0.91 & \textbf{0.91} & 0.90 \\ \cmidrule{2-8}
 & Un-Krum & \textbf{0.91} & 0.92 & \textbf{0.92} & 0.90 & 0.89 & \textbf{0.90} \\ \cmidrule{2-8}
 & Un-Med & 0.94 & \textbf{0.95} & \textbf{0.95} & 0.93 & \textbf{0.91} & 0.90 \\ \bottomrule
\end{tabular}
\end{adjustbox}
\label{tab:DLs_untar}
\end{table*}

\begin{table*}[!b]
\centering
\caption{Performance comparison of the PenTiDef module on the CIC-IDS2018 and Edge-IIoTset datasets with two DL models, CNN and SqueezeNet (untargeted attack)}
\begin{adjustbox}{width=\linewidth}
\begin{tabular}{ccccccccc}
\toprule
\multirow{2}{*}{\textbf{Dataset}} & \multirow{2}{*}{\textbf{Attack}} & \multicolumn{2}{c}{\textbf{10\% adv}} & \multicolumn{2}{c}{\textbf{20\% adv}} & \multicolumn{2}{c} {\textbf{40\% adv}} \\ \cmidrule{3-8} 
 &  & \textbf{CNN} & \textbf{SqueezeNet} & \textbf{CNN} & \textbf{SqueezeNet} & \textbf{CNN} & \textbf{SqueezeNet} \\ \toprule
\multicolumn{8}{c}{\textbf{IID}} \\ \toprule
\multirow{4}{*}{CIC-IDS2018} & GAN-SL & \textbf{0.97} & 0.96 & 0.97 & 0.97 & 0.94 & \textbf{0.95} \\ \cmidrule{2-8} 
 & GAN-ML & \textbf{0.96} & 0.95 & 0.96 & \textbf{0.97} & \textbf{0.97} & 0.96 \\ \cmidrule{2-8}
 & GAN-Con & \textbf{0.97} & 0.96 & 0.97 & \textbf{0.98} & 0.95 & \textbf{0.96} \\ \cmidrule{2-8}
 & BD & \textbf{0.97} & 0.96 & \textbf{0.97} & 0.96 & 0.94 & \textbf{0.95} \\ \midrule
\multirow{4}{*}{Edge-IIoTset} & GAN-SL & 0.92 & \textbf{0.93} & 0.93 & \textbf{0.94} & \textbf{0.95} & 0.94 \\ \cmidrule{2-8}
 & GAN-ML & 0.93 & \textbf{0.94} & 0.93 & \textbf{0.94} & \textbf{0.95} & 0.94 \\ \cmidrule{2-8}
 & GAN-Con & 0.93 & \textbf{0.95} & \textbf{0.95} & 0.94 & \textbf{0.95} & 0.94 \\ \cmidrule{2-8}
 & BD & 0.95 & \textbf{0.97} & \textbf{0.97} & 0.96 & \textbf{0.95} & 0.94 \\ \bottomrule
\multicolumn{8}{c}{\textbf{non-IID}} \\ \toprule
\multirow{4}{*}{CIC-IDS2018} & GAN-SL & 0.92 & \textbf{0.93} & 0.94 & \textbf{0.93} & \textbf{0.92} & 0.93 \\ \cmidrule{2-8} 
 & GAN-ML & \textbf{0.92} & 0.91 & \textbf{0.92} & 0.91 & 0.92 & \textbf{0.94} \\ \cmidrule{2-8}
 & GAN-Con & \textbf{0.95} & 0.93 & \textbf{0.93} & 0.94 & \textbf{0.95} & 0.94 \\ \cmidrule{2-8}
 & BD & \textbf{0.95} & 0.93 & \textbf{0.93} & 0.94 & \textbf{0.95} & 0.94 \\ \midrule
\multirow{4}{*}{Edge-IIoTset} & GAN-SL & \textbf{0.89} & 0.88 & 0.89 & \textbf{0.90} & 0.89 & \textbf{0.90} \\ \cmidrule{2-8}
 & GAN-ML & \textbf{0.91} & 0.90 & \textbf{0.91} & 0.90 & 0.91 & \textbf{0.92} \\ \cmidrule{2-8}
 & GAN-Con & \textbf{0.93} & 0.92 & 0.92 & \textbf{0.93} & \textbf{0.93} & 0.90 \\ \cmidrule{2-8}
 & BD & \textbf{0.95} & 0.94 & \textbf{0.95} & 0.94 & 0.95 & \textbf{0.96} \\ \bottomrule
\end{tabular}
\end{adjustbox}
\label{tab:DLs_tar}
\end{table*}


\textbf{Table \ref{tab:Targeted_non_iid}} shows us that PenTiDef remains the model with the most stable performance compared to the other two attack methods, although all three modules experience a decrease in accuracy due to the influence of the non-IID dataset. Specifically, PenTiDef often achieves the highest or near-highest values in most attacks, such as LF, WS, Untargeted-Krum, and Untargeted-Med. This demonstrates PenTiDef's good capability in distinguishing between benign and malicious models. Even as the ratio of attackers increases, PenTiDef maintains high performance, showing the stability and effectiveness of this method in protecting the system from poisoning attacks. Compared to other methods, PenTiDef always shows a clear difference in metrics. PenTiDef effectively preserves this consistency by ensuring that benign models retain high CKA scores while isolating adversarial updates. In contrast, FedCC struggles to maintain stable CKA scores, particularly in non-IID environments. The uneven data distribution introduces variations in local model updates, leading to inconsistencies that FedCC fails to manage effectively. This lack of stability causes difficulties in distinguishing between benign and malicious models, ultimately reducing FedCC’s detection accuracy. 

In summary, PenTiDef stands out as a robust and reliable defense method in protecting distributed ML systems from poisoning attacks. Similar to the statistics in the IID case, targeted attacks in the non-IID case also slightly reduce the performance of all three modules. As seen in  Figure \ref{fig:cka_non_iid}, due to the impact of the uneven data distribution, the CKA scores for the benign clients have decreased (by approximately 0.1). The CKA scores of FedCC show a lack of stability and clarity in classifying malicious and benign models. This explains why the PenTiDef model has higher performance in most scenarios compared to FedCC.

Finally, to address the question of cost efficiency, \textbf{Table \ref{tab:training_time}} summarizes the training time of each defense model. It can be seen that our PenTiDef model has a significantly faster training time compared to the other two models, namely FedCC and FLARE. This can be explained as follows:
\begin{itemize}
    \item Firstly, the FLARE model uses an algorithm to compute the distance between two PLR vectors. Due to the complexity of this algorithm, FLARE has the slowest training time compared to the other two defense models.
    \item Secondly, the FedCC model operates more similarly to our model than FLARE. However, FedCC calculates similarity based on PLRs using CKA scores, whereas our model computes it based on LSRs extracted from PLRs through an AE.
\end{itemize}
Therefore, our model achieves the fastest training time, addressing the issue of resource cost and enhancing scalability in distributed FL environments.

\begin{table*}[!b]
\centering
\caption{Benchmark results of the Hyperledger Fabric network with 5000 transactions}
\begin{adjustbox}{width=\textwidth}
    \begin{tabular}{ccccccccc}
\toprule
\multicolumn{8}{c}{\textbf{5 transactions/s}} \\ \toprule
Name & Succ & Fail & Send Rate (TPS) & Max Latency (s) & Min Latency (s) & Avg Latency (s)  & Throughput (TPS)  \\ \hline
CreateModel & 5000 & 0 & 28.1 & 2.10 & 0.03 & 0.21 & 28.0 \\ \midrule
QueryAllModels & 5000 & 0 & 64.1 & 0.29 & 0.04 & 0.10 & 64.1  \\ \midrule
QueryLastModel & 5000 & 0 & 127.0 & 0.28 & 0.02 & 0.05 & 126.9  \\ \midrule
QueryModelsBySelectedClientID & 5000 & 0 & 114.4 & 0.25 & 0.02 & 0.06 & 114.3  \\ \midrule
QueryModelsByModelIndex & 5000 & 0 & 97.8 & 0.34 & 0.02 & 0.06 & 97.7 \\ \midrule
QueryModelsByAdversaryClientID & 5000 & 0 & 86.4 & 0.49 & 0.02 & 0.07 & 86.4 \\ \midrule
QueryModelsByBenignClientID & 5000 & 0 & 78.9 & 0.25 & 0.02 & 0.08 & 78.9  \\ \toprule

\multicolumn{8}{c}{\textbf{20 transactions/s}} \\ \toprule
Name & Succ & Fail & Send Rate (TPS) & Max Latency (s) & Min Latency (s) & Avg Latency (s)  & Throughput (TPS)  \\ \midrule
CreateModel & 5000 & 0 & 86.5 & 2.09 & 0.03 & 0.12 & 84.5  \\ \midrule
QueryAllModels & 5000 & 0 & 82.1 & 0.32 & 0.04 & 0.17 & 82.1 \\ \midrule
QueryLastModel & 5000 & 0 & 158.8 & 0.27 & 0.02 & 0.10 & 158.6  \\ \midrule
QueryModelsBySelectedClientID & 5000 & 0 & 151.1 & 0.28 & 0.02 & 0.10 & 151.0  \\ \midrule
QueryModelsByModelIndex & 5000 & 0 & 177.7 & 0.22 & 0.02 & 0.09 & 177.5 \\ \midrule
QueryModelsByAdversaryClientID & 5000 & 0 & 129.8 & 0.53 & 0.02 & 0.10 & 129.6 \\ \midrule
QueryModelsByBenignClientID & 5000 & 0 & 113.9 & 0.46 & 0.02 & 0.12 & 113.8  \\ \bottomrule
\end{tabular}
\end{adjustbox}
\label{tab:5000}
\end{table*}

\subsection{Result 3 (Answer to Scenario 3): PenTiDef demonstrates model-agnostic robustness across CNN and SqueezeNet architectures}
For this scenario, we conduct experiments on the attack scenarios to test the performance of the PenTiDef module. This time, we conduct the experiments on two DL models, including CNN and SqueezeNet. This aims to evaluate the detection and defense capabilities of PenTiDef on different neural network architectures, thereby determining the stability and effectiveness of the module in various conditions and environments. We conduct experiments on both IID and non-IID cases with (\textbf{Table \ref{tab:DLs_untar}}) and targeted attacks (\textbf{Table \ref{tab:DLs_tar}}). Similar to scenario 3, the model performs most stably in detecting untargeted attacks on the IID dataset and shows the most fluctuation in the non-IID dataset with targeted attacks.

Based on the statistics tables, the CNN model often achieves higher or equal results compared to SqueezeNet in many different attack scenarios. This indicates that CNN tends to perform more effectively than SqueezeNet in the tested scenarios. However, both models have good detection and defense capabilities, but the stability and effectiveness of the model can be influenced by the type of attack and the data structure used.

Notably, our PenTiDef module has demonstrated stability and flexibility by performing well across various DL models. This shows that PenTiDef not only has strong detection and defense capabilities but can also be flexibly applied to different neural network architectures. The ability to adapt to different models and maintain stable performance proves that PenTiDef is a comprehensive and effective defense solution in protecting DFL-IDS from potential attacks.

\begin{table*}[!t]
\centering
\caption{Benchmark results of the Hyperledger Fabric network with 10000 transactions}
\begin{adjustbox}{width=\textwidth}
\begin{tabular}{ccccccccc}
\toprule
\multicolumn{8}{c}{\textbf{5 transactions/s}} \\ \toprule
Name & Succ & Fail & Send Rate (TPS) & Max Latency (s) & Min Latency (s) & Avg Latency (s)  & Throughput (TPS)  \\ \midrule
CreateModel & 10000 & 0 & 27.6 & 2.09 & 0.02 & 0.22 & 27.5 \\ \midrule
QueryAllModels & 10000 & 0 & 32.8 & 0.45 & 0.07 & 0.18 & 32.8 \\ \midrule
QueryLastModel & 10000 & 0 & 69.9 & 0.25 & 0.03 & 0.09 & 69.9 \\ \midrule
QueryModelsBySelectedClientID & 10000 & 0 & 57.2 & 0.36 & 0.04 & 0.12 & 57.2 \\ \midrule
QueryModelsByModelIndex & 10000 & 0 & 67.7 & 0.25 & 0.03 & 0.10 & 67.7 \\ \midrule
QueryModelsByAdversaryClientID & 10000 & 0 & 47.0 & 0.50 & 0.03 & 0.15 & 47.0  \\ \midrule
QueryModelsByBenignClientID & 10000 & 0 & 48.9 & 0.43 & 0.04 & 0.13 & 48.9 \\ \bottomrule

\multicolumn{8}{c}{\textbf{20 transactions/s}} \\ \toprule
Name & Succ & Fail & Send Rate (TPS) & Max Latency (s) & Min Latency (s) & Avg Latency (s)  & Throughput (TPS)  \\ \midrule
CreateModel & 10000 & 0 & 33.1 & 2.07 & 0.03 & 0.11 & 33.7 \\ \midrule
QueryAllModels & 10000 & 0 & 39.4 & 0.67 & 0.07 & 0.35 & 39.4 \\ \midrule
QueryLastModel & 10000 & 0 & 91.5 & 0.45 & 0.04 & 0.16 & 91.5 \\ \midrule
QueryModelsBySelectedClientID & 10000 & 0 & 43.9 & 1.23 & 0.05 & 0.36 & 43.9 \\ \midrule
QueryModelsByModelIndex & 10000 & 0 & 71.4 & 0.55 & 0.03 & 0.21 & 71.4 \\ \midrule
QueryModelsByAdversaryClientID & 10000 & 0 & 45.9 & 0.97 & 0.05 & 0.34 & 45.9 \\ \midrule
QueryModelsByBenignClientID & 10000 & 0 & 47.4 & 1.00 & 0.06 & 0.31 & 47.4  \\ \bottomrule
\end{tabular}
\end{adjustbox}
\label{tab:10000}
\end{table*}

\subsection{Result 4 (Answer to Scenario 4): The blockchain coordination layer remains efficient and stable under increasing transaction loads} 

In each evaluation, we executed all the tasks defined in the chaincode to interact with the system, including: CreateModel, GetAllModels, GetLastModel, QueryModelsBySelectedClientID, QueryModelsByModelIndex, QueryModelsByAdversaryClientID, QueryModelsByBenignClientID. Transactions are sent continuously, increasing from 5 to 20 transactions per second, to the system for evaluation.

Tables \ref{tab:5000} and \ref{tab:10000} present the benchmark results of the Hyperledger Fabric network with 5000 and 10000 transactions. In terms of stability, both tables show the network's stability, with all transactions being successful. As the number of transactions increased from 5000 to 10000, the sending rate tended to decrease, especially at the rate of 5 transactions per second. The average and maximum latency slightly increased with the higher number of transactions, likely due to the heavier load on the network. The throughput remained relatively high and stable, demonstrating the good performance of the Hyperledger Fabric network in handling transactions. Overall, the Hyperledger Fabric network showed good performance at both load levels of 5000 and 10000 transactions, with all transactions being successful and high throughput.

\section{Discussion}
\label{sec:discus}

The experimental results demonstrate that PenTiDef achieves a superior balance among privacy protection, robustness against poisoning attacks, and decentralization for Federated Intrusion Detection Systems (FL-IDS) in heterogeneous Industrial IoT (IIoT) environments. By consistently outperforming FLARE and FedCC across both IID and realistic non-IID settings on CIC-IDS2018 and Edge-IIoTSet, with adversary ratios up to 40\%, PenTiDef delivers higher detection accuracy and F1-score against a wide spectrum of poisoning attacks, including label-flipping, weight-scaling, Krum/Med variants, backdoor, and GAN-based strategies, while incurring lower overall training overhead. These improvements arise directly from the synergistic integration of client-side Distributed Differential Privacy (DDP), latent-space anomaly detection (AE-compressed LSRs with CKA clustering), and blockchain-orchestrated validation, which together enable secure FedAvg aggregation without requiring auxiliary datasets, prior knowledge of adversary numbers, or a central server.

From a practical perspective, PenTiDef addresses critical deployment barriers in multi-site IIoT infrastructures. The elimination of centralized aggregation mitigates single points of failure and enhances fault tolerance across administrative domains. The immutable ledger and smart-contract enforcement provide transparent auditability and incentive-compatible participation, features particularly valuable for collaborative cybersecurity among untrusted industrial partners. Moreover, the stochastic DDP mechanism with bounded noise ($\sigma \in [0,0.2]$) delivers meaningful $(\epsilon,\delta)$-guarantees while preserving model utility, as evidenced by only marginal convergence degradation. The latent-space module further stabilizes representation learning under severe non-IID distributions—common in heterogeneous sensors, protocols, and production workflows—thereby improving generalizability beyond what centralized or semi-decentralized baselines can achieve.

Despite these advancements, several limitations warrant acknowledgment. First, evaluations are currently limited to binary classification tasks. While effective for distinguishing benign versus malicious traffic, extending PenTiDef to multi-class or multi-label attack categorization is necessary for fine-grained threat intelligence in real-world IIoT deployments. Second, although the framework demonstrates strong resilience against the tested poisoning strategies, its performance against highly adaptive or stealthy attacks—such as dynamic backdoors, model inversion combined with poisoning, or collusion among malicious nodes—remains to be fully characterized. Third, the blockchain layer, while lightweight in the current 6-node Fabric setup, introduces additional communication and computational overhead that may become non-negligible at larger scales (hundreds of edge devices) or under constrained IIoT hardware. Finally, the fixed noise bound in DDP, while practical, creates a potential trade-off surface that could be exploited by free-rider or noise-injection adversaries if not dynamically adjusted based on observed trust scores.

Future research directions include: (i) adaptive noise scheduling mechanisms that leverage latent-space trust feedback to optimize the privacy-utility trade-off on-the-fly; (ii) integration of explainable AI (XAI) techniques to interpret CKA-based decisions and improve operator trust; (iii) large-scale experiments with real IIoT testbeds (e.g., using ns-3 or physical edge devices) to assess energy consumption, straggler tolerance, and cross-domain generalization; and (iv) formal verification of smart-contract security properties and composition theorems for the end-to-end privacy guarantees. A comprehensive cost-benefit analysis encompassing transaction throughput, latency under varying loads, and resource footprints on resource-constrained gateways would further guide industrial adoption.

In conclusion, PenTiDef represents a significant step toward practical, trustworthy decentralized FL-IDS by holistically addressing the intertwined challenges of privacy leakage, poisoning vulnerability, and centralized coordination risks in heterogeneous IIoT ecosystems. With continued refinement and real-world validation, the framework holds strong potential to enable secure collaborative intrusion detection across critical industrial infrastructures.

\section{Conclusion and Future Directions}\label{sec:conclusion}

This paper presented PenTiDef, a fully decentralized, privacy-preserving, and poisoning-resilient framework for Federated Intrusion Detection Systems tailored to heterogeneous Industrial IoT (IIoT) environments. By synergistically integrating client-side Distributed Differential Privacy (DDP) with stochastic Gaussian noise, a latent-space anomaly detection module based on AutoEncoder-compressed Latent Semantic Representations (LSRs) and Centered Kernel Alignment (CKA) clustering, and a blockchain-orchestrated coordination layer built on Hyperledger Fabric, PenTiDef effectively addresses three fundamental challenges in decentralized FL-IDS: gradient leakage, malicious model updates, and centralized single points of failure.

Extensive experiments on CIC-IDS2018 and Edge-IIoTSet under both IID and realistic non-IID data distributions demonstrate that PenTiDef consistently outperforms state-of-the-art defenses such as FLARE and FedCC. The framework achieves higher detection accuracy and F1-score across diverse poisoning attacks (label-flipping, weight-scaling, Krum/Med, backdoor, and GAN-based) with adversary ratios up to 40\%, while maintaining competitive training efficiency and only marginal utility loss from DDP perturbation. These results validate the effectiveness of the proposed unified secure aggregation protocol, which operates without auxiliary datasets, prior adversary knowledge, or a trusted central server. While PenTiDef advances the state of the art in decentralized and privacy-preserving FL-IDS, certain limitations remain. The current evaluation focuses on binary classification, and its resilience against highly adaptive or collusive attacks requires further characterization. In addition, the computational and communication overhead of the blockchain layer at very large scales needs deeper investigation, particularly on resource-constrained IIoT devices.

Future work will extend PenTiDef to multi-class and multi-label intrusion detection, incorporate adaptive noise scheduling and explainable AI techniques to further optimize the privacy-utility trade-off, and conduct large-scale evaluations on physical IIoT testbeds. Formal verification of the smart-contract logic and comprehensive cost-benefit analyses will also be pursued to facilitate industrial adoption. In conclusion, PenTiDef provides a practical and scalable foundation for secure collaborative intrusion detection in adversarial, heterogeneous IIoT ecosystems. By holistically addressing privacy, robustness, and decentralization, the framework offers a promising pathway toward trustworthy federated learning solutions for critical cyber-physical systems, including smart manufacturing, energy grids, and other distributed infrastructures.






\bibliography{references}

\end{document}